\documentclass[oldversion]{aa}  

\usepackage{graphicx}
\usepackage{txfonts}
\usepackage{natbib}
\usepackage{supertabular}
\usepackage{longtable}
\usepackage{url}
\usepackage{dcolumn}
\usepackage{placeins}
\usepackage{multirow}
\begin{document}

\newcommand{\zabs}{\ensuremath{z_{\rm abs}}}
\newcommand{\zem}{\ensuremath{z_{\rm em}}}
\newcommand{\HH}{\mbox{H$_2$}}
\newcommand{\HD}{\mbox{HD}}
\newcommand{\DD}{\mbox{D$_2$}}
\newcommand{\CO}{\mbox{CO}}
\newcommand{\dla}{damped Lyman-$\alpha$}
\newcommand{\Dla}{Damped Lyman-$\alpha$}
\newcommand{\lya}{Ly-$\alpha$}
\newcommand{\lyb}{Ly-$\beta$}
\newcommand{\lyg}{Ly-$\gamma$}
\newcommand{\ArI}{\ion{Ar}{i}}
\newcommand{\CaII}{\ion{Ca}{ii}}
\newcommand{\CI}{\ion{C}{i}}
\newcommand{\CII}{\ion{C}{ii}}
\newcommand{\CIV}{\ion{C}{iv}}
\newcommand{\ClI}{\ion{Cl}{i}}
\newcommand{\ClII}{\ion{Cl}{ii}}
\newcommand{\CrII}{\ion{Cr}{ii}}
\newcommand{\CuII}{\ion{Cu}{ii}}
\newcommand{\DI}{\ion{D}{i}}
\newcommand{\FeI}{\ion{Fe}{i}}
\newcommand{\FeII}{\ion{Fe}{ii}}
\newcommand{\HI}{\ion{H}{i}}
\newcommand{\MgI}{\ion{Mg}{i}}
\newcommand{\MgII}{\ion{Mg}{ii}}
\newcommand{\MnII}{\ion{Mn}{ii}}
\newcommand{\NI}{\ion{N}{i}}
\newcommand{\NV}{\ion{N}{v}}
\newcommand{\NiII}{\ion{Ni}{ii}}
\newcommand{\OI}{\ion{O}{i}}
\newcommand{\OVI}{\ion{O}{vi}}
\newcommand{\PII}{\ion{P}{ii}}
\newcommand{\PbII}{\ion{Pb}{ii}}
\newcommand{\SI}{\ion{S}{i}}
\newcommand{\SII}{\ion{S}{ii}}
\newcommand{\SiII}{\ion{Si}{ii}}
\newcommand{\SiIV}{\ion{Si}{iv}}
\newcommand{\TiII}{\ion{Ti}{ii}}
\newcommand{\ZnII}{\ion{Zn}{ii}}
\newcommand{\AlII}{\ion{Al}{ii}}
\newcommand{\AlIII}{\ion{Al}{iii}}

\newcommand{\Ho}{\mbox{H$_0$}}
\newcommand{\angstrom}{\mbox{{\rm \AA}}}
\newcommand{\abs}[1]{\left| #1 \right|} 
\newcommand{\avg}[1]{\left< #1 \right>} 
\newcommand{\kms}{\ensuremath{{\rm km\,s^{-1}}}}
\newcommand{\cmsq}{\ensuremath{{\rm cm}^{-2}}}
\newcommand{\thisqso}{SDSS\,J160457$+$220300}
\newcommand{\thisqsolong}{SDSS\,J160457.50$+$220300.5}
\newcommand{\thisqsoshort}{Q\,1604$+$2203}


\title{Diffuse molecular gas at high redshift
\thanks{Based on observations carried out with the 
Ultraviolet and Visual Echelle Spectrograph, mounted on the ESO Very Large Telescope, under 
Prgm.~ID.~081.A-0334(B).}}

\subtitle{Detection of CO molecules and the 2175~{\AA} dust feature at $z=1.64$}


\author{P. Noterdaeme\inst{1,3}, C. Ledoux\inst{2}, R. Srianand\inst{3}, P. Petitjean\inst{1} and S. Lopez\inst{4}
}
\authorrunning{P. Noterdaeme et al.}
\titlerunning{Diffuse molecular gas at high redshift}

   \institute{
Universit\'e Paris 6, Institut d'Astrophysique de Paris, CNRS UMR 
7095, 98bis bd Arago, 75014 Paris, France - \email{[noterdaeme, petitjean]@iap.fr}
\and
European Southern Observatory, Alonso de C\'ordova 3107, Casilla 19001, Vitacura,
 Santiago 19, Chile - \email{cledoux@eso.org}
\and
Inter-University Centre for Astronomy and Astrophysics, Post Bag 4, Ganeshkhind, 411\,007 Pune, India
 - \email{[pasquiern, anand]@iucaa.ernet.in}
\and
Departamento de Astronom\'ia, Universidad de Chile, Casilla 36-D, Santiago, Chile - \email{slopez@das.uchile.cl}
             }

\offprints{P. Noterdaeme}

\date{}

\abstract{
We present the detection of carbon monoxide molecules (CO) at $z=1.6408$ towards the quasar \thisqsolong\ 
using the Very Large Telescope Ultraviolet and Visual Echelle Spectrograph. 
CO absorption is detected in at least two components in the first six A-X bands and one 
d-X(5-0) inter-band system. This is the second detection of this kind along a quasar line of sight. 
The CO absorption profiles are well modelled assuming a rotational excitation of CO 
in the range 6~$<$~$T_{\rm ex}$~$<$~16~K, which is consistent with or higher than the temperature of the 
Cosmic Microwave Background Radiation at this redshift.
We derive a total CO column density of $N($CO$)=4\times10^{14}~\cmsq$. 
The measured column densities of  \SI, \MgI, \ZnII, \FeII\ and 
\SiII\ indicate a dust depletion pattern typical of cold gas in the Galactic disc.  
The background quasar spectrum is significantly reddened (u$-$K~$\sim4.5$~mag) 
and presents a pronounced 
2175~$\angstrom$ dust absorption feature at the redshift of the CO absorber. 
Using a control sample of $\sim$500 quasars we find the chance probability for this feature
to be spurious is $\sim$0.3\%. We show that the 
spectral energy distribution (SED) of the quasar is well fitted with a QSO composite spectrum 
reddened with a Large Magellanic Cloud supershell extinction law at the redshift of the absorber. 
It is noticeable that this quasar is absent from the colour-selected SDSS quasar sample. This demonstrates our 
current view of the Universe may be biased against dusty sightlines.
These direct observations of carbonaceous molecules and dust open up the possibility of studying physical 
conditions and chemistry of diffuse molecular gas in high redshift galaxies.
}

\maketitle

\section{Introduction}

Quasar absorption lines provide a powerful tool to detect and study gaseous baryonic 
matter at all redshifts in a luminosity-unbiased way. 
Large column densities of neutral gas are revealed by the damped Lyman-$\alpha$ absorption lines 
they imprint in the spectrum of background quasars.
Because of the large neutral hydrogen column densities ($N(\HI)\ga 10^{20}~\cmsq$), 
similar to what is observed along Galactic lines of sight, and the presence of metals at 
different levels of chemical enrichment \citep[e.g.][]{Pettini97, Prochaska02}, 
it is believed that a large fraction of Damped Lyman-$\alpha$ systems (DLAs) are 
located close to regions of star formation at high redshift 
\citep[see, e.g.][for a review on the subject]{Wolfe05}. 

The typical dust-to-gas ratio in DLAs, is generally less 
than one tenth of what is observed in the local interstellar medium (ISM) 
and only a small fraction ($\sim$10-15\%) of them show detectable amounts of 
molecular hydrogen \citep{Ledoux03,Noterdaeme08}. 
Even in these cases, the molecular fractions are small compared to what is seen in the Galactic ISM. 
It is therefore likely that most of the DLAs probe only diffuse neutral gas \citep{Petitjean00}. 
In contrast, the detection of the cold ($T\sim 10-100$~K), dusty and molecular gas, 
a fundamental ingredient for star-formation, is elusive till now in absorption studies. 
The corresponding regions could have been missed so far due to the large extinction 
they are expected to produce.  
It could also be that molecular gas escapes detection because of its very small 
cross section \citep{Zwaan06}. 

Gamma Ray Bursts (GRBs) are thought to be located in star forming regions, and the 
associated absorptions are therefore more likely to arise from the densest part of 
the ISM at high redshift. However, the physical state of the absorbing gas is very 
likely to be influenced by the intense UV radiation field arising from the GRB itself 
\citep[e.g.][]{Vreeswijk07}. This could explain the absence of H$_2$ in the majority 
of GRB-DLAs \citep{Tumlinson07}. However, only a handful of GRB-DLAs have been studied 
so far and existing data are still consistent with the statistics of H$_2$ detections 
in intervening QSO-DLA samples (Ledoux et al. 2009, submitted). When molecules are seen 
\citep{Prochaska09}, their excitation is observed to be high, indicating strong UV 
pumping from the GRB afterglow. 

We have started a programme to search for cold gas along QSO lines of sight. 
The huge number of quasar spectra available in the Sloan Digital Sky Survey (SDSS) 
and the faint magnitude limit achieved by the survey allows us to identify 
absorption systems having unique characteristics. In particular the
systems can be selected on the basis of the presence of \CI\ which should
flag predominantly cold neutral gas. 
We searched for strong ($W_{r}(\CI\lambda1656)\sim 0.5$~{\AA}) \CI\ 
absorbers at $z\sim1.5-3$ along the line of sight of $\sim$ 40\,000 QSOs from 
the SDSS Data Release 7.

The selection lead to the first detection of carbon monoxide (CO) absorption 
lines at $z\sim2.4$ \citep{Srianand08}.
Similarly, selection of strong \MgII\ systems at intermediate redshift lead
to the detection of the 2175~\angstrom\ feature and 21-cm absorption in two 
$z\sim1.3$ \MgII\ systems \citep{Srianand08bump}.
Till now, direct signatures of dust at $z>0$, such as the UV bump or diffuse interstellar bands 
(DIBs) have only been reported in a small number of cases \citep[e.g.,][]{Motta02, Wucknitz03, 
Ellison08b, Liang09}.

In this paper we present the simultaneous detection of carbon monoxide absorption lines 
and 2175~\angstrom\ dust feature in an intervening absorber at $z=1.64$ towards \thisqso. 
We present high spectral resolution observations in Sect.~2, the metal content of the system in Sect.~3 and the 
analysis of molecular lines in Sect.~4. We discuss the extinction and presence of a UV bump 
in Sect.~5. Finally, we conclude in Sect.~6.

\section{Observations}

The quasar \thisqsolong\ ($\zem=1.98$) was observed with the Ultraviolet and Visual Echelle 
Spectrograph \citep[UVES;][]{Dekker00} in visitor mode on June 28, 29, and 30, 2008.
The total exposure time on source is 29\,400~s.  
Both UVES spectrographic arms were used 
simultaneously taking advantage of a dichroic setting with central wavelengths of 390~nm in the blue and 564~nm 
in the red. The resulting wavelength coverage is 329-451~nm and 462-665~nm with a gap between 
559 and 568~nm corresponding to the physical gap between the two red CCDs. The CCD pixels 
were binned $2\times2$ and the slit width adjusted to 1$\arcsec$ to match the ambient seeing conditions. 
This yielded a resolving power of R~=~47\,500 in the blue and R~=~45\,000 in the red as measured from Th-Ar lines 
from the calibration lamp. 
The data were reduced using the MIDAS-based UVES pipeline v\,2.9.7, which 
performs an accurate tracking of the object while subtracting the sky spectrum at the same time. Cosmic ray 
impacts and CCD defects were rejected iteratively. 
Wavelengths were rebinned to the vacuum-heliocentric rest frame and individual exposures were co-added using 
a sliding windows and weighting the signal by the signal-to-noise ratio in each pixel. 
We analysed the spectrum using standard Voigt-profile fitting techniques.
Oscillator strengths and wavelengths of CO absorption lines were taken from \citet{Morton94} with updated 
values from \citet{Eidelsberg03} for the inter-band systems. Heavy element abundances are given 
relative to solar \citep{Grevesse07}, with [X/H$]\equiv\log N($X$)/N($H$)-\log$~(X/H)$_\odot$ and 
assuming $N($H$)=N(\HI)$.
For short reference through the paper, we refer to \thisqsolong\ as \thisqsoshort. 
\section{Metal content \label{secmetals}}

\begin{figure}
\centering
\begin{tabular}{cc}
\includegraphics[bb=219 244 393 623,clip=,angle=90,width=0.46\hsize]{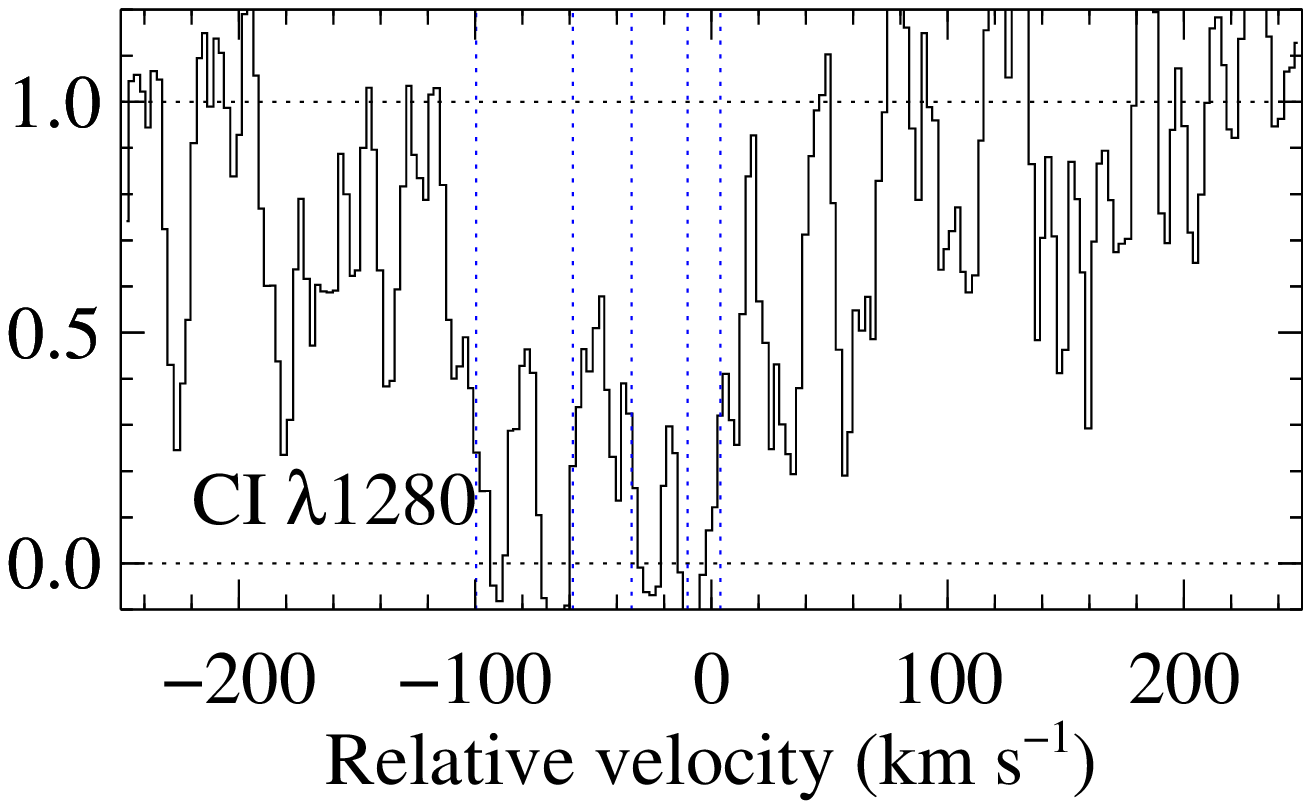}    &\includegraphics[bb=219 244 393 623,clip=,angle=90,width=0.46\hsize]{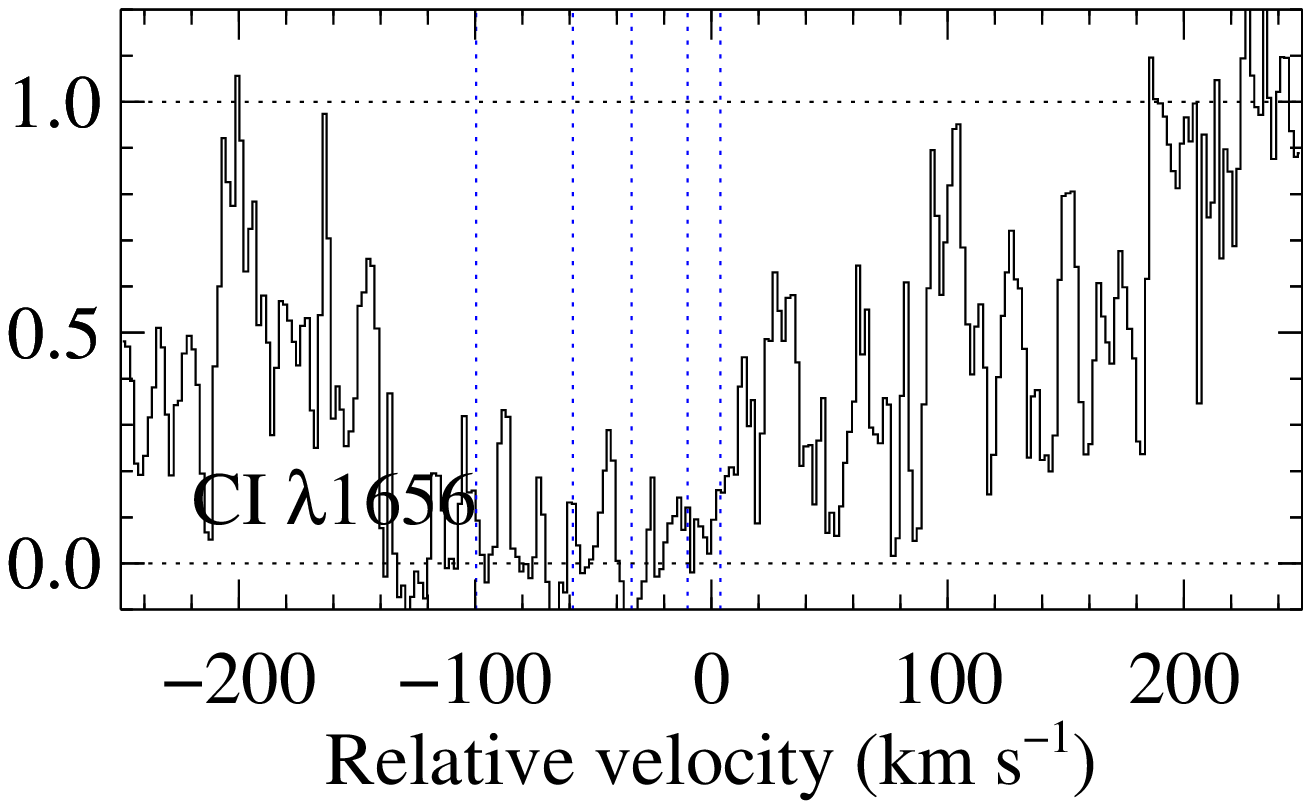}    \\
\includegraphics[bb=219 244 393 623,clip=,angle=90,width=0.46\hsize]{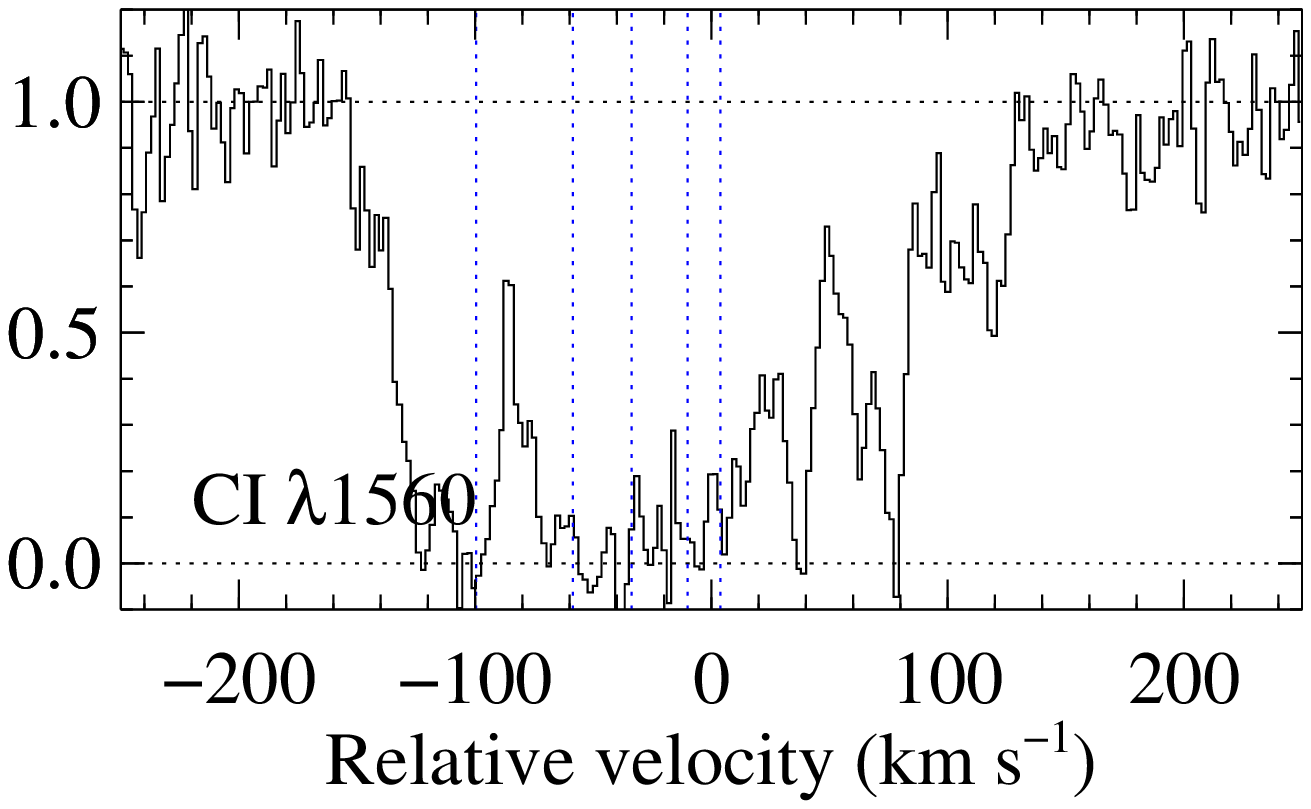}    &\includegraphics[bb=219 244 393 623,clip=,angle=90,width=0.46\hsize]{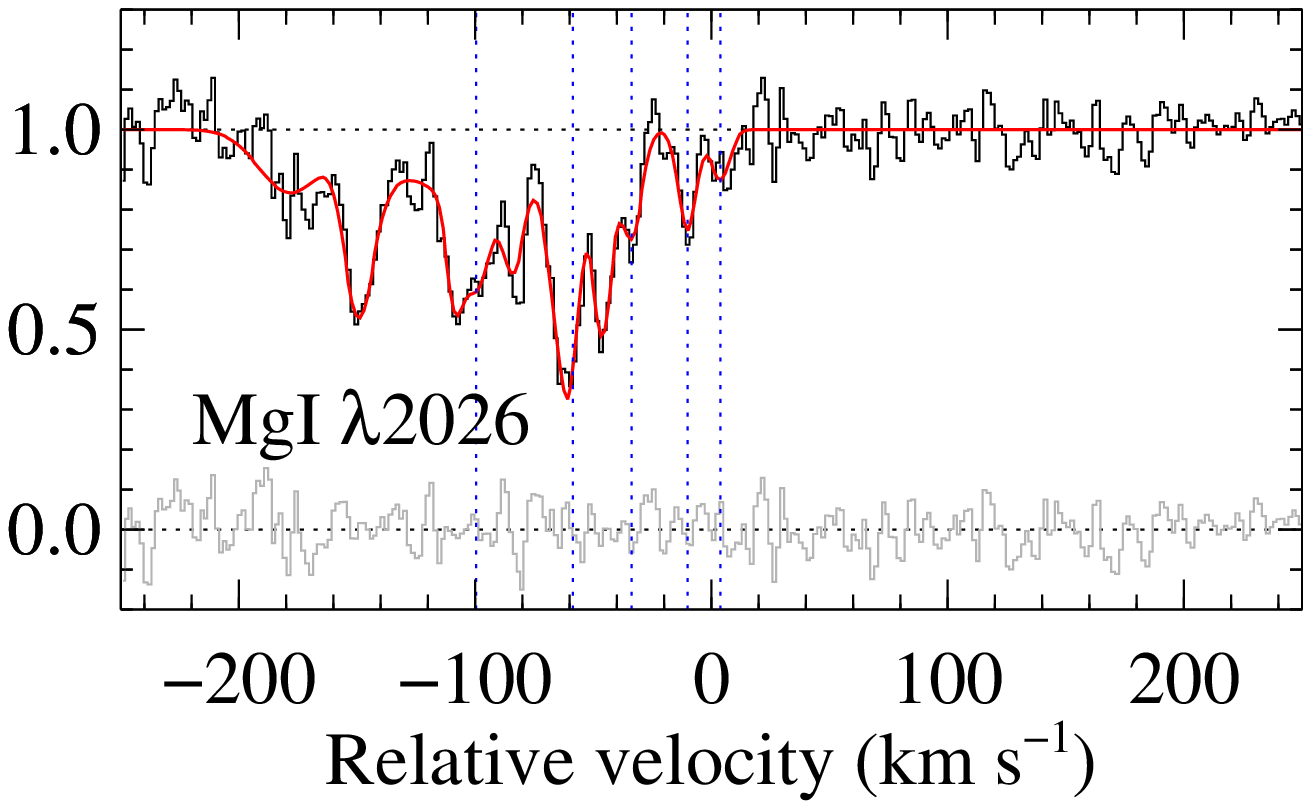}    \\
\includegraphics[bb=219 244 393 623,clip=,angle=90,width=0.46\hsize]{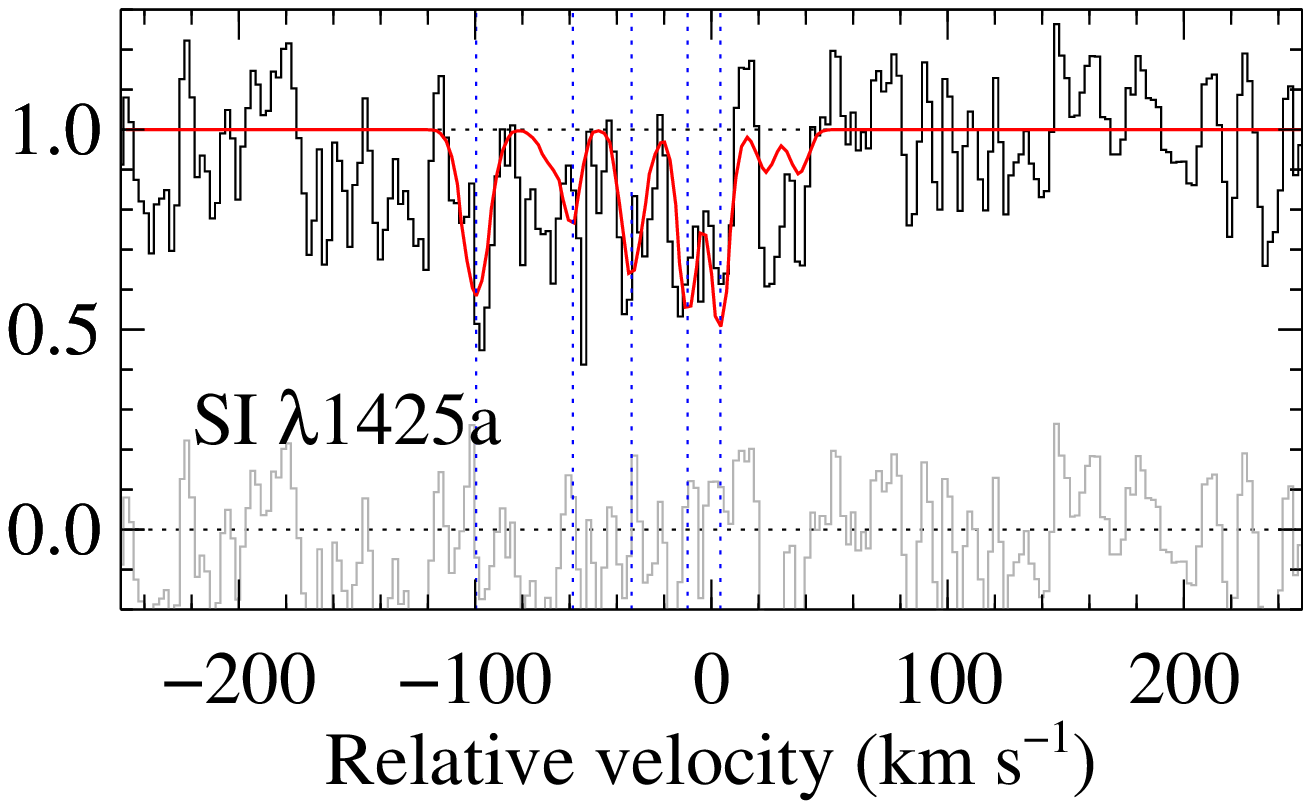}   &\includegraphics[bb=219 244 393 623,clip=,angle=90,width=0.46\hsize]{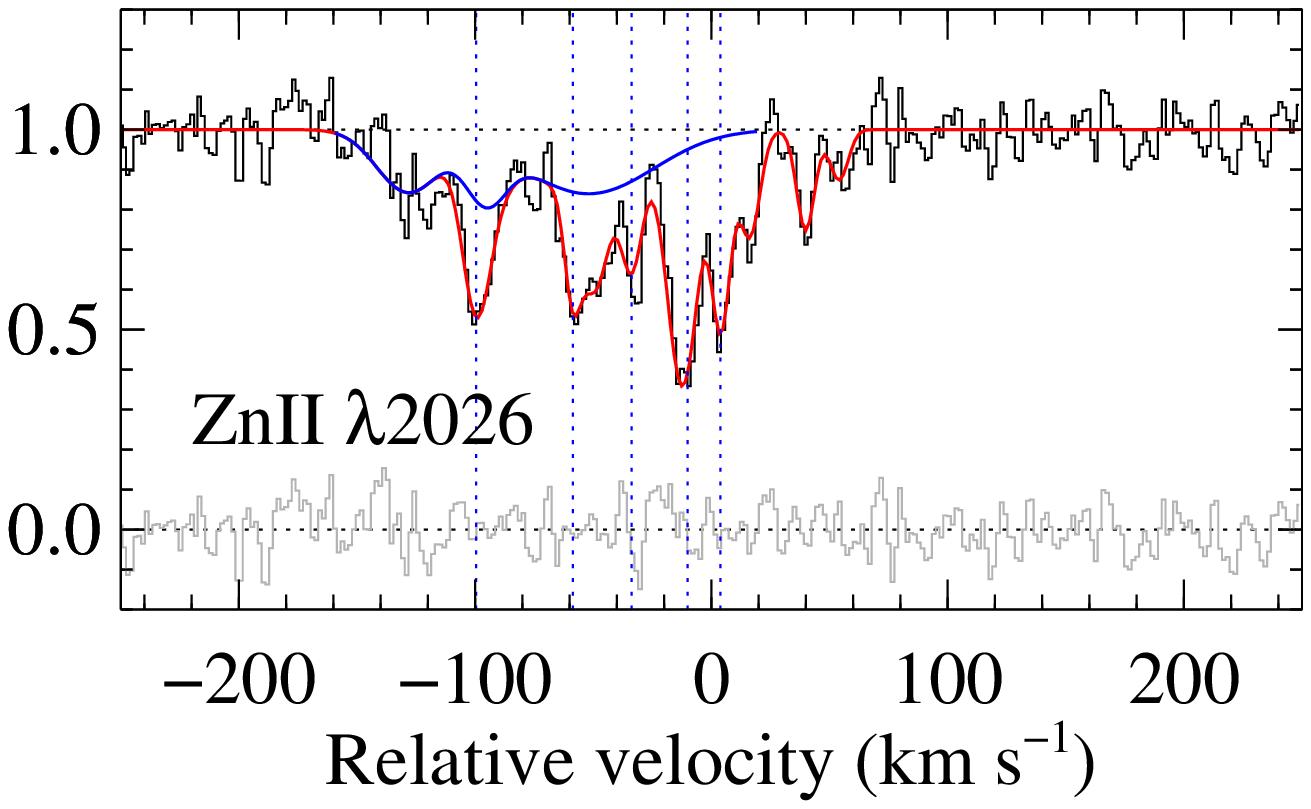}\\ 
\includegraphics[bb=219 244 393 623,clip=,angle=90,width=0.46\hsize]{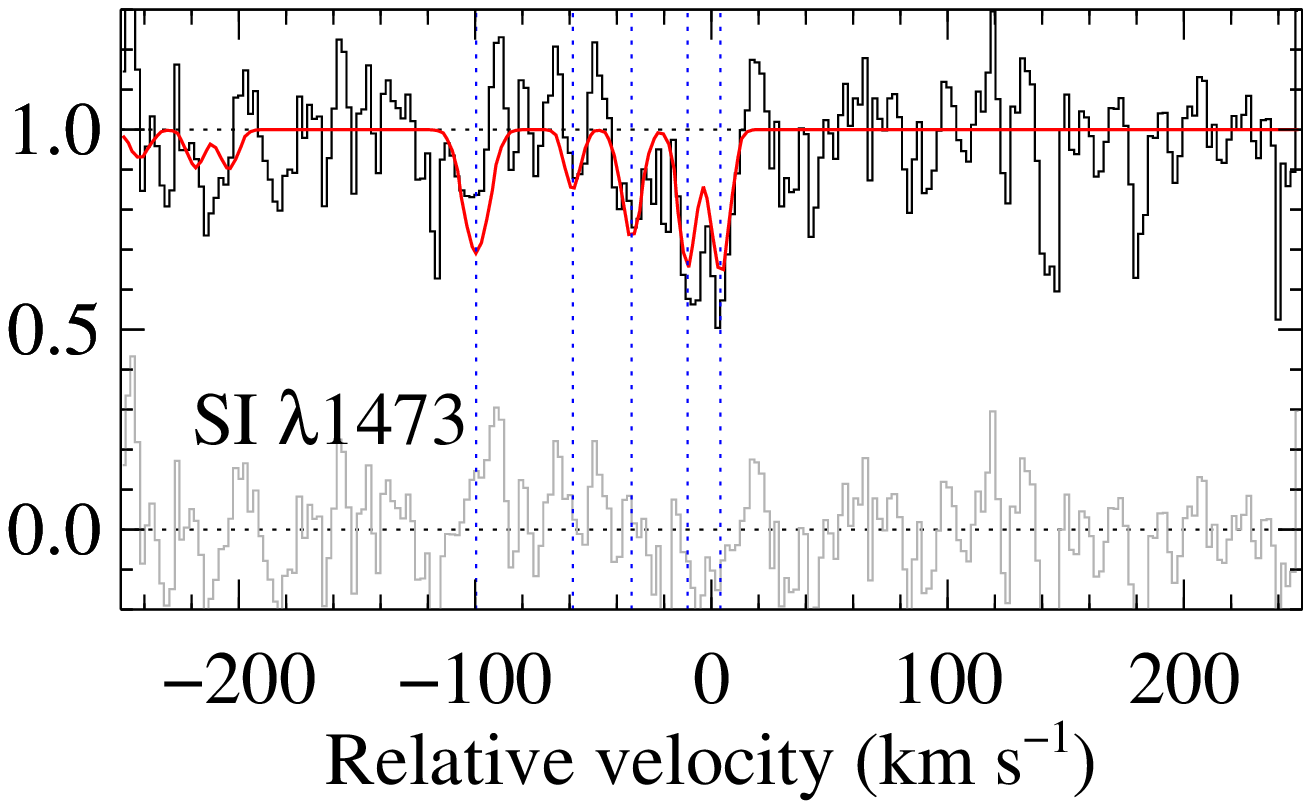}    &\includegraphics[bb=219 244 393 623,clip=,angle=90,width=0.46\hsize]{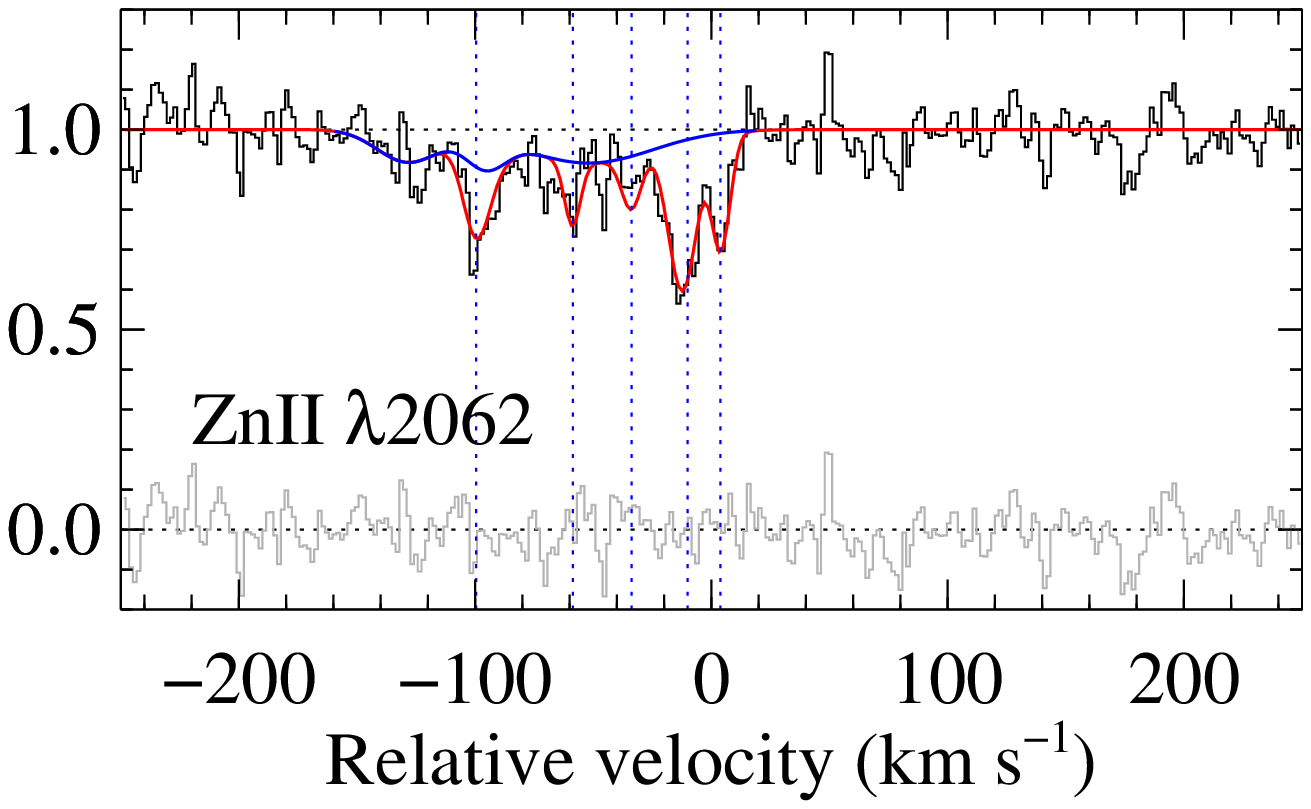}\\ 
\includegraphics[bb=219 244 393 623,clip=,angle=90,width=0.46\hsize]{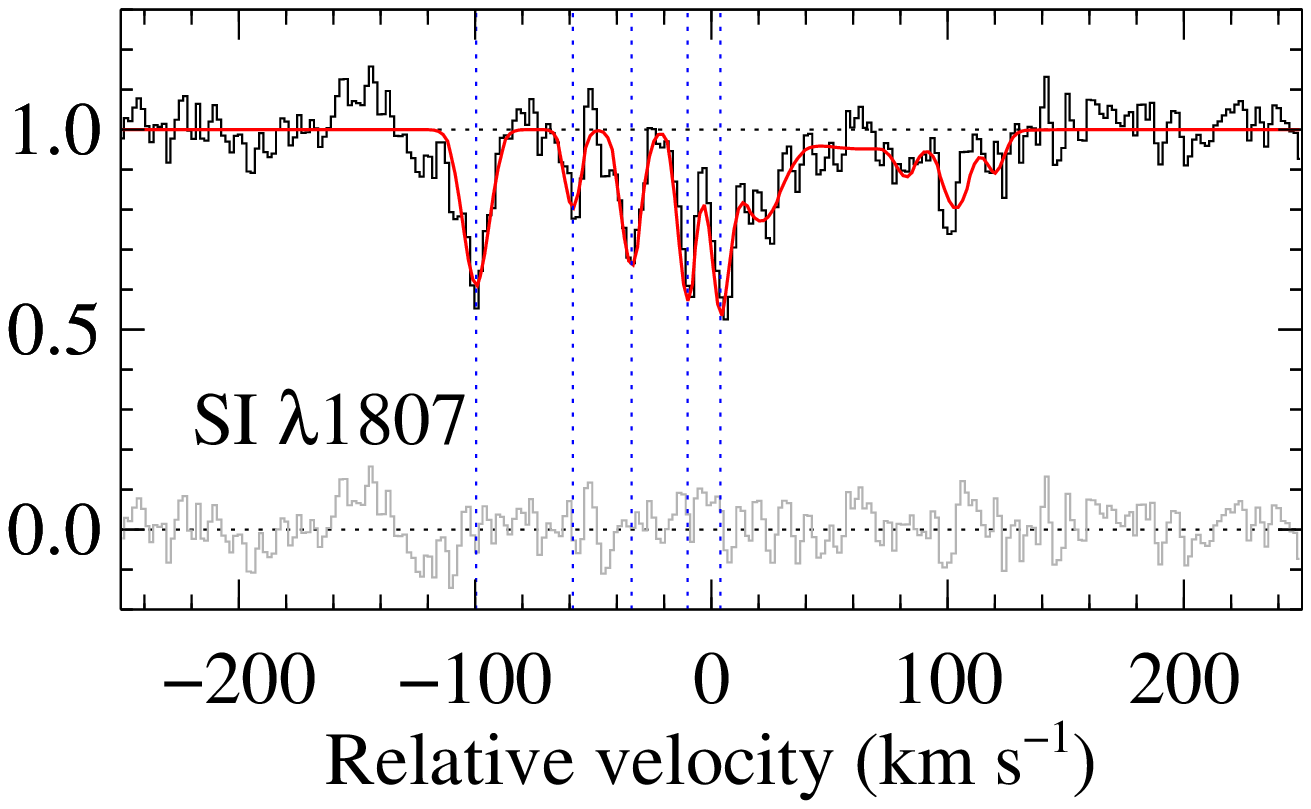}    &\includegraphics[bb=219 244 393 623,clip=,angle=90,width=0.46\hsize]{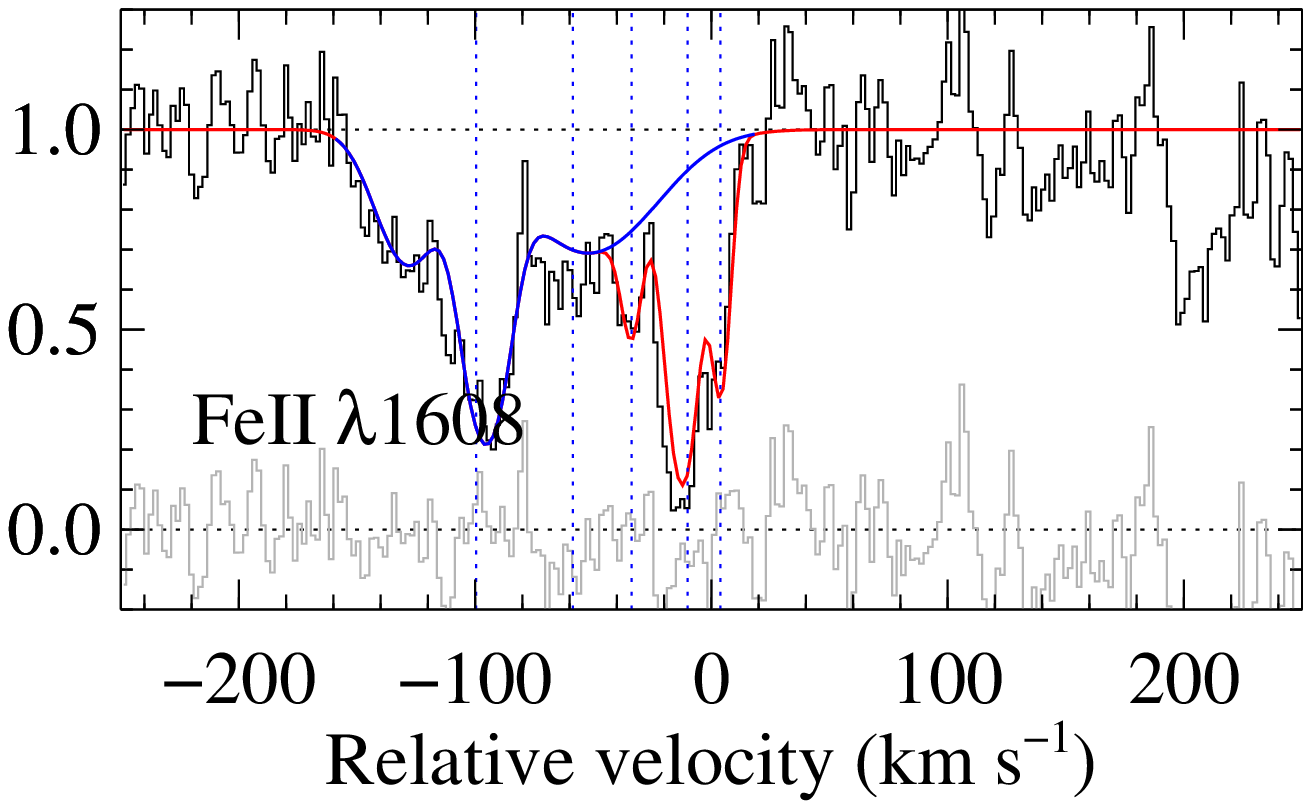}\\ 
\includegraphics[bb=165 244 393 623,clip=,angle=90,width=0.46\hsize]{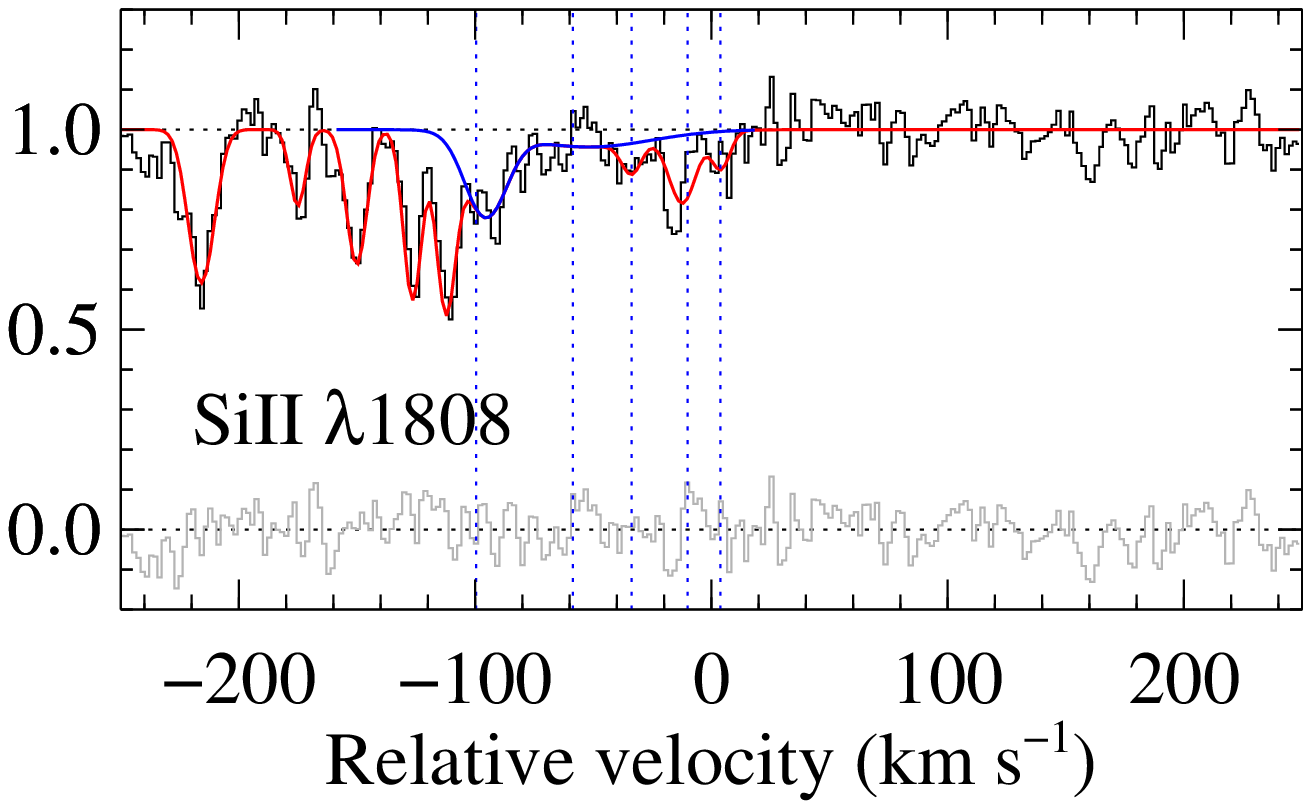}&\includegraphics[bb=165 244 393 623,clip=,angle=90,width=0.46\hsize]{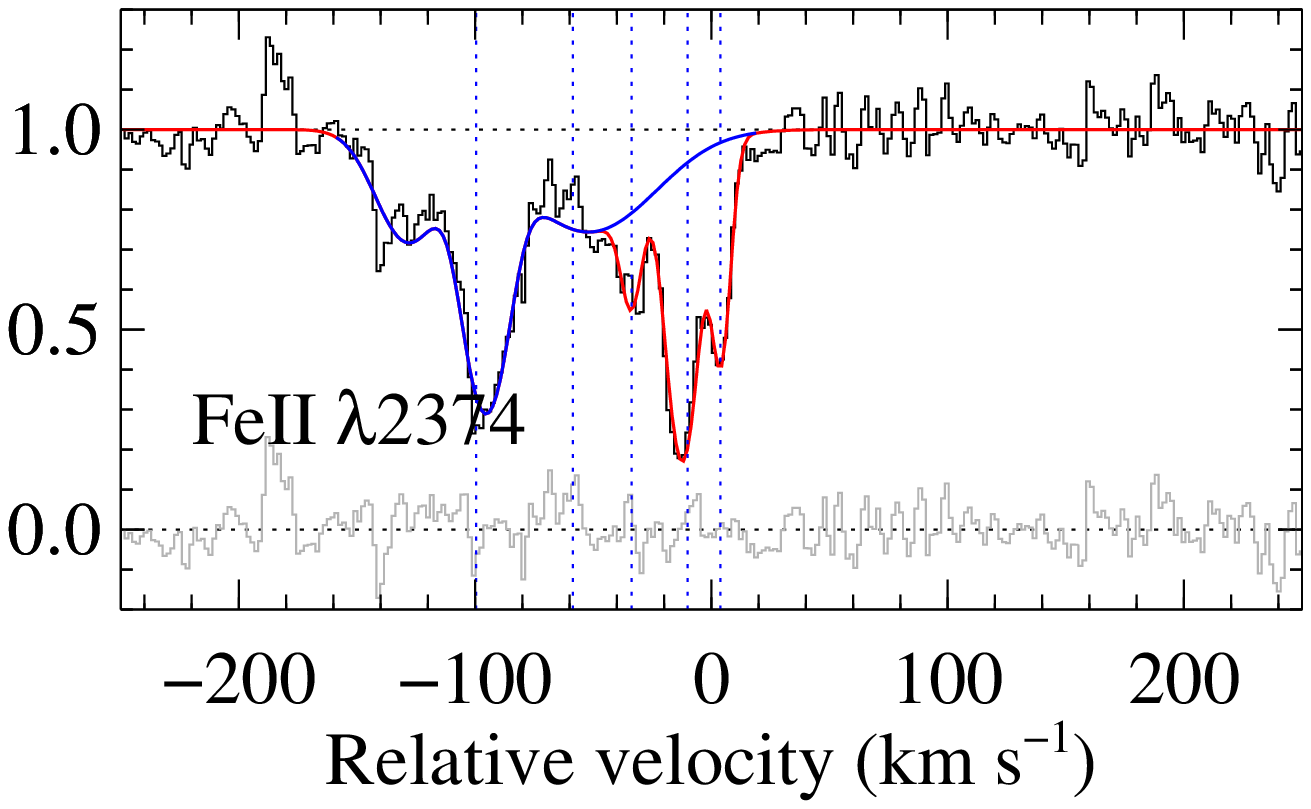}\\                                                                                                        
\end{tabular}
\caption{\label{metals} Voigt-profile fit to neutral and singly ionised species. The origin of 
the velocity scale is taken at $z=1.6408$. Vertical dotted lines indicate the position of narrow components ('N' in Table~\ref{table}). 
The contribution from the broad components only ('B' in Table~\ref{table}) is represented by the blue profile.
The extremely strong \CI\ lines, though not fitted, are also shown in the top panels. The \CI\,$\lambda$1280 profile has been 
smoothed by 2 pixels for presentation purpose only.}
\end{figure}

Absorption lines from singly ionised metals (\FeII, \SiII, and \ZnII) 
as well as neutral species (\SI, \MgI, and \CI) are detected around $\zabs=1.6405$ over $\sim$200~\kms
(see Fig.~\ref{metals}). The covered absorption lines of neutral species are exceptionally strong, with 
\CI\,$\lambda\lambda$1560,1656 being heavily saturated and \SI\ clearly detected 
in several transitions. Note that \SI\ is rarely seen in QSO absorbing systems \citep{Quast08,Srianand08}.
\FeI\ is not detected down to $\log N(\FeI)=11.4$ for each component (3\,$\sigma$ upper limit).
We performed simultaneous multicomponent Voigt-profile fits with {\sl fitlyman} \citep{Fontana95} to 
constrain redshifts, Doppler parameters $b$ and column densities, see Table~\ref{table}. 
Five narrow components are required to fit the neutral species (\SI\ and \MgI). 
The profiles of singly ionised species (\ZnII, \SiII\ and \FeII) require additional broad components.

While the narrow components can still be seen in the profiles 
of \ZnII\ transitions, the presence of additional broad components introduces 
a degeneracy in the results in particular the relative column densities in the
narrow and broad components. We have somewhat artificially associated the broad and 
narrow components
so that one could consider the sum of the column densities in the 
broad ($b>10~\kms$) and narrow ($b<10~\kms$) components to be representative of the column 
density in the six components listed in Table~1.
Indeed, narrow components can be completely lost into the 
broad ones as it is the case for \FeII\ and \SiII. 
{\sl However, because lines are in the optically-thin regime, integrated column densities 
will not depend upon the actual number of components}.
The above decomposition shows that the coldest gas, as traced by the S~{\sc i} components,
is found in clumps embedded in a more turbulent and probably warmer phase.
This structure is natural \citep[see e.g.][]{Petitjean92} and is usually observed
in the H$_2$ phase of DLAs \citep[][]{Ledoux02}.

The immediate consequence of this is that the depletion factor estimated from integrated
column densities may not be a correct representation of the actual depletion factors 
in individual components. 

Using the integrated column densities (Table~\ref{table}) and upper limits
of different ions ($\log N(\NiII)<13.2$ and $\log N(\CrII)<12.8$ at 
3\,$\sigma$ c.l., for the whole profile) 
we find, [Fe/Zn]~$=-1.47$, [Si/Zn]~$=-1.07$, [Ni/Zn]~$<-1.5$ and [Cr/Zn]~$<-1.3$. 
This is consistent with the depletion pattern seen in the
cold neutral medium of the Galactic ISM and that of the Large Magellanic Cloud \citep[see][]{Welty99}.

From the equivalent width of \CI\,$\lambda$1280, which is the weakest \CI\ line 
available we derive $\log N(\CI)>15.3$, in the linear regime limit. However, even 
this line is saturated, and filling the C\,{\sc i} profiles with five components with the same
parameters  -- Doppler parameters and redshifts --  as the \SI\ components gives 
$\log N(\CI)>16.0$ which is still a lower-limit. 
Therefore, this system has by far the highest column densities of \CI\ and \SI\ 
known in any QSO absorption system. 

\begin{table*}
\caption{Results of Voigt-profile fitting to detected lines at $z\sim1.64$ in the spectrum of \thisqsoshort\
\label{table}}
\begin{center}
\begin{tabular}{c c r c c c c c c}
\hline
\hline
\multirow{2}{*}{Comp.$^a$}\rule[0pt]{0pt}{10pt} & \multirow{2}{*}{$z$}       & \multicolumn{1}{c}{\multirow{2}{*}{$b$ (\kms)}}                     & \multicolumn{6}{c}{$\log N(X)$ (\cmsq)}\\
                 & &                      &   \ZnII               & \FeII\                & \SiII\                & \MgI\                 & \SI\              & CO\\
  \hline
1B    & 1.63967	& 17.8	$\pm$ 0.8	& 12.31	$\pm$ 0.03	& 13.73	$\pm$ 0.02	& 	      	        & 	      	        & 	            &       \\
 & & & & & & & & \\                                                                                  
2N    & 1.63992	& 6.3	$\pm$ 0.4	& 12.34	$\pm$ 0.05	& 	      	        & 	      	        & 12.85	$\pm$ 0.04	& 13.18	$\pm$ 0.02  & $\le$13.64 \\
2B    & 1.63996	& 11.4	$\pm$ 0.3	& 12.09	$\pm$ 0.11	& 14.10	$\pm$ 0.01	& 14.72	$\pm$ 0.03	& 	      	        & 	            &       \\
 & & & & & & & & \\                                                                                  
3N    & 1.64028	& 2.8	$\pm$ 0.6	& 12.07	$\pm$ 0.04	& 	      	        & 	      	        & 	      	        & 12.64	$\pm$ 0.05  & $\le$13.28 \\
3B    & 1.64034	& 37.6	$\pm$ 1.4	& 12.67	$\pm$ 0.04	& 14.00	$\pm$ 0.02	& 14.52	$\pm$ 0.09	& 	      	        & 	            &       \\
 & & & & & & & & \\                                                                              
4N    & 1.64080	& 4.5	$\pm$ 0.4	& 12.06	$\pm$ 0.05	& 13.29	$\pm$ 0.04	& 13.96	$\pm$ 0.13	& 12.73	$\pm$ 0.03	& 13.01	$\pm$ 0.02  & $\le$13.80 \\
 & & & & & & & & \\                                                                               
\multirow{2}{*}{5N}    & 1.64069	& 6.8	$\pm$ 0.3	& 12.70	$\pm$ 0.03	& 14.08	$\pm$ 0.01	& 14.45	$\pm$ 0.05	& 	      	        & 	            &       \\
                       & 1.64071	& 3.6	$\pm$ 0.4	&                	&             	        & 	      	        & 12.66	$\pm$ 0.04	& 13.10	$\pm$ 0.02  & 14.24 $\pm$ 0.20\\
 & & & & & & & & \\                                                                                     
6N    & 1.64083	& 3.9	$\pm$ 0.3	& 12.42	$\pm$ 0.02	& 13.67	$\pm$ 0.02	& 14.02	$\pm$ 0.10	& 12.33	$\pm$ 0.07	& 13.13	$\pm$ 0.02  & 14.34 $\pm$ 0.10\\
 & & & & & & & & \\                                                                               
total &         &                       & 13.30 $\pm$ 0.02      & 14.67 $\pm$ 0.01      & 15.13 $\pm$ 0.03      & 13.28 $\pm$ 0.02      & 13.75 $\pm$ 0.01  & 14.59 $\pm$ 0.11 $^b$\\
\hline
\end{tabular}
\end{center}
\footnotesize
$^a$ Narrow components ($b<10~\kms$) are indicated by ``N'' while ``B'' stands for broad component ($b>10~\kms$).\\
$^b$ Total $N($CO$)$ measured from comp. 5 and 6 only. Undetected components (2N, 3N and 4N) 
    could increase this value by up to 0.1~dex (see text).\\
\normalsize
\end{table*}

\section{Carbon monoxide}

Carbon monoxide absorption lines are detected in several \mbox{A-X} bands and the \mbox{d-X\,(5-0)} 
inter-band system.

\begin{figure}
\centering
\begin{tabular}{cc}
\includegraphics[bb=219 244 393 623,clip=,angle=90,width=0.46\hsize]{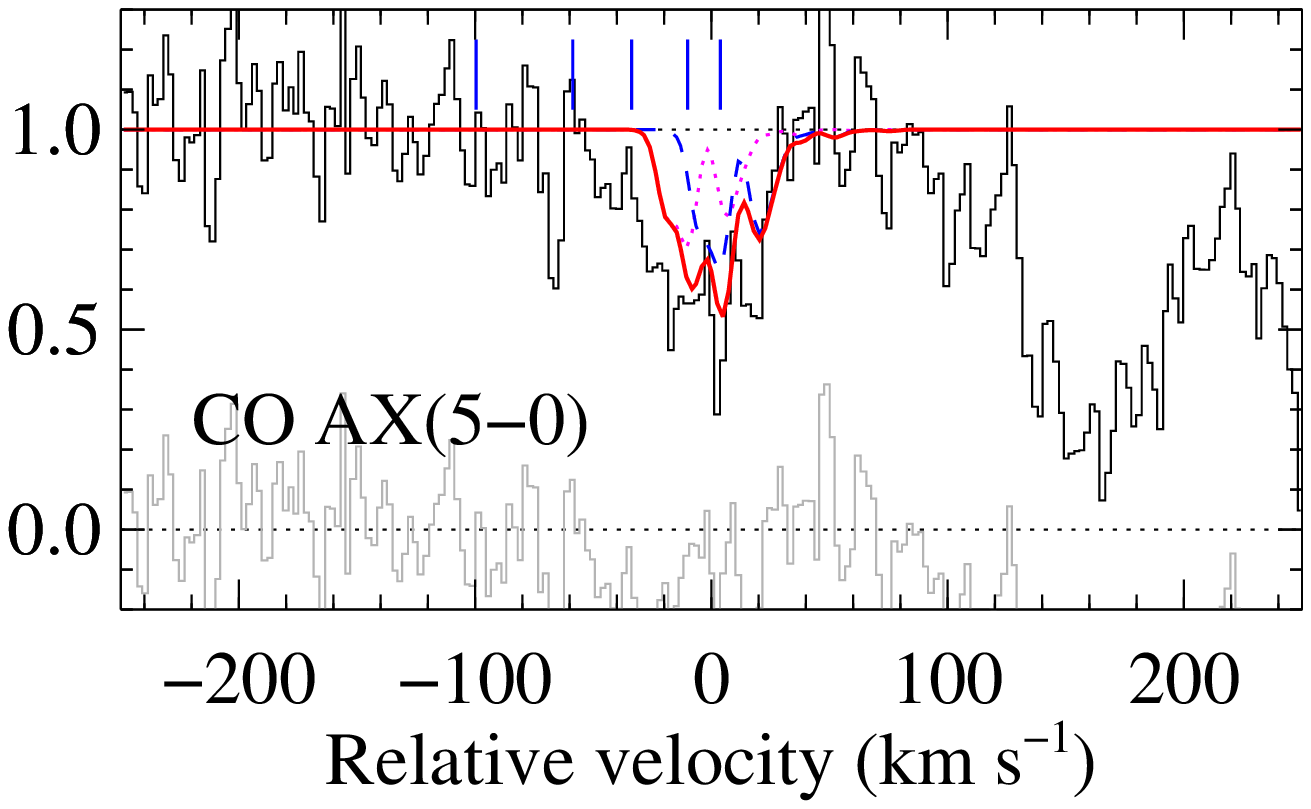}&
\includegraphics[bb=219 244 393 623,clip=,angle=90,width=0.46\hsize]{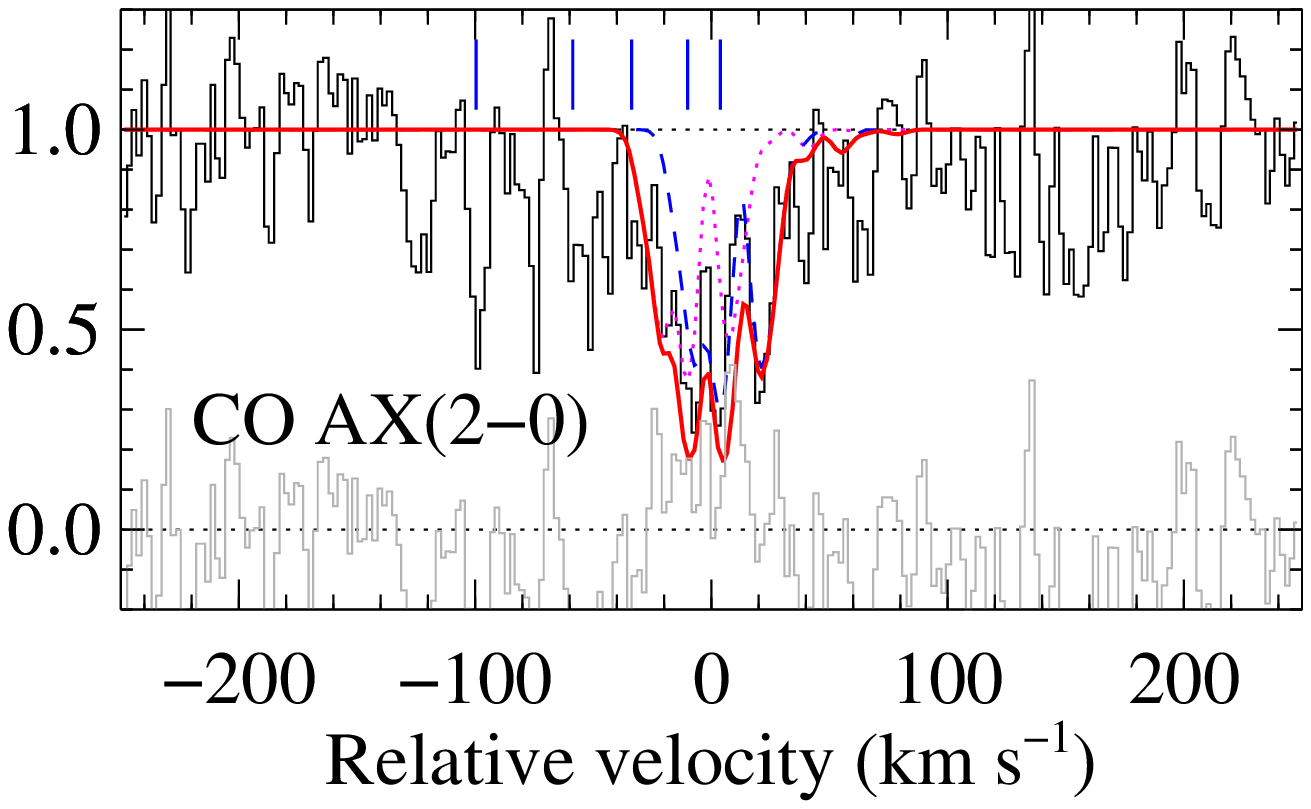}\\
\includegraphics[bb=219 244 393 623,clip=,angle=90,width=0.46\hsize]{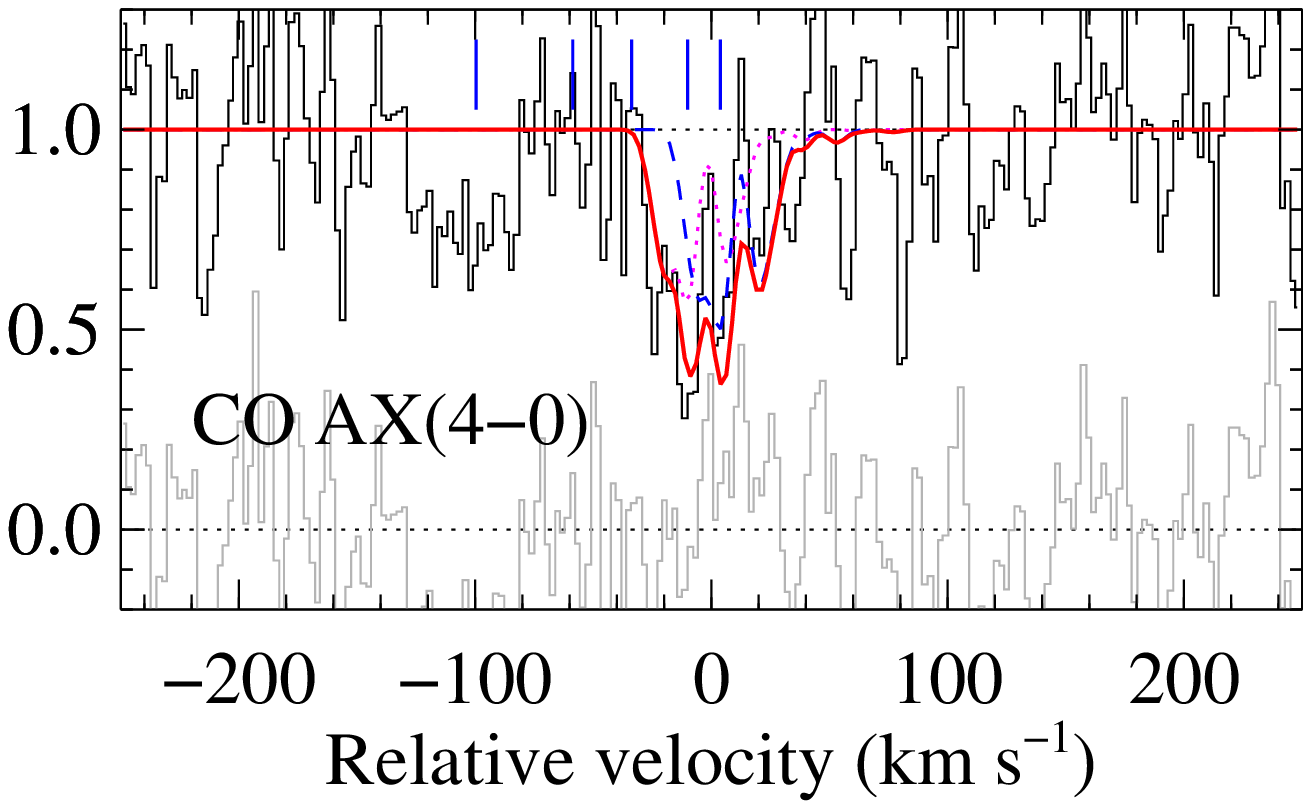}&
\includegraphics[bb=219 244 393 623,clip=,angle=90,width=0.46\hsize]{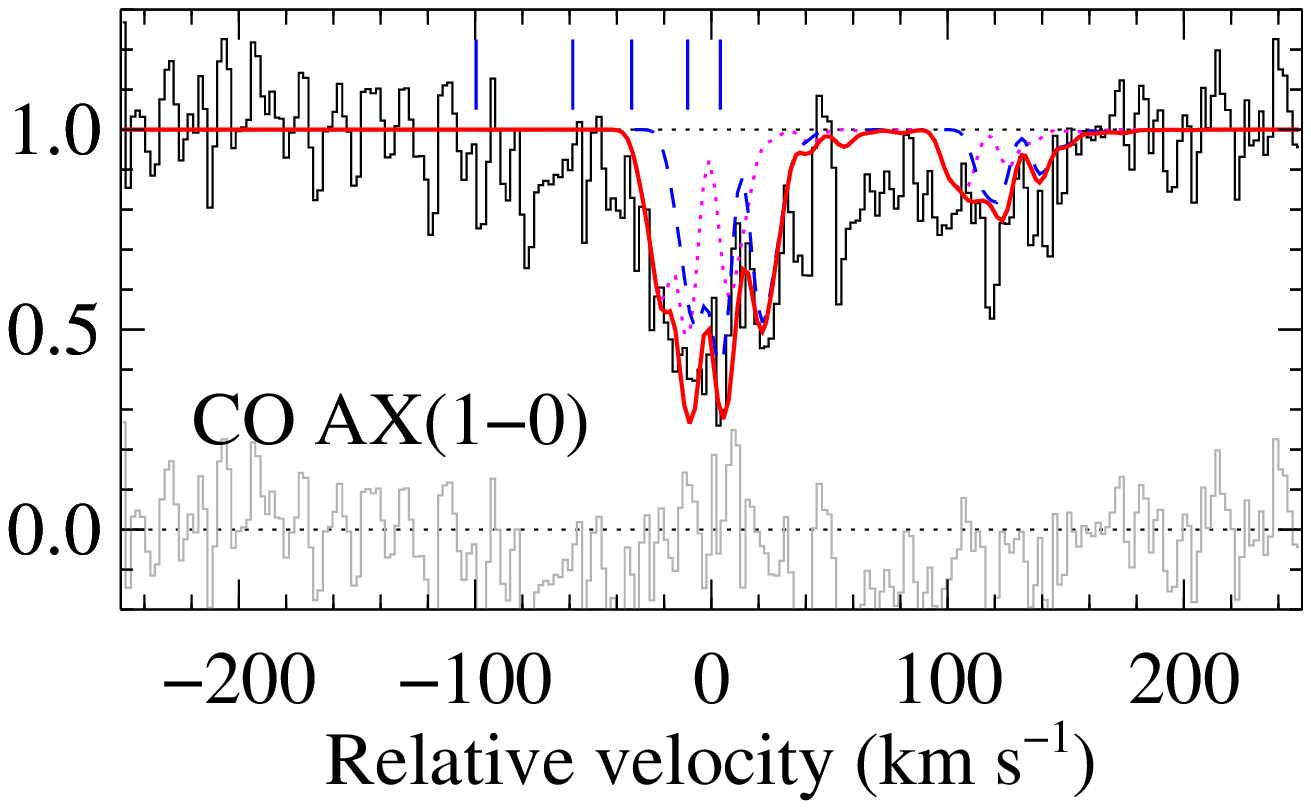}\\ 
\includegraphics[bb=165 244 393 623,clip=,angle=90,width=0.46\hsize]{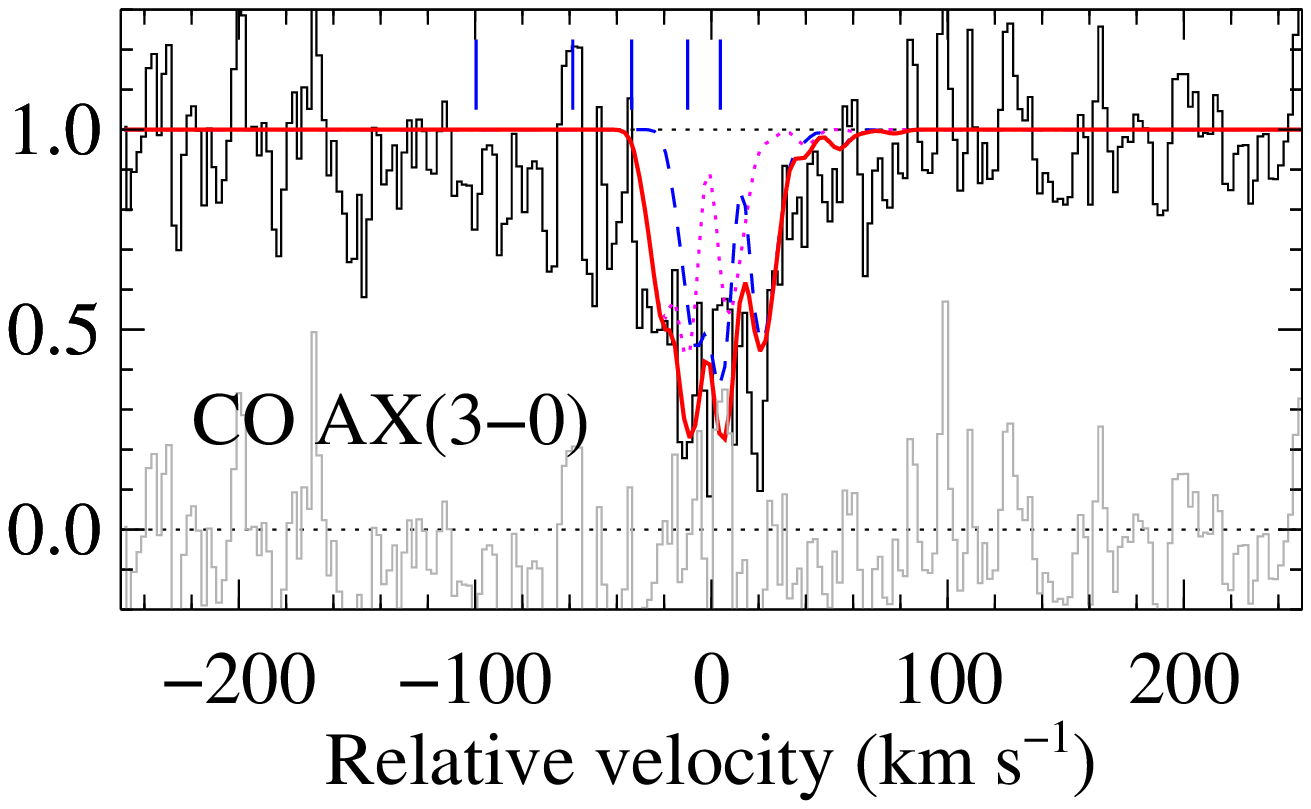}&
\includegraphics[bb=165 244 393 623,clip=,angle=90,width=0.46\hsize]{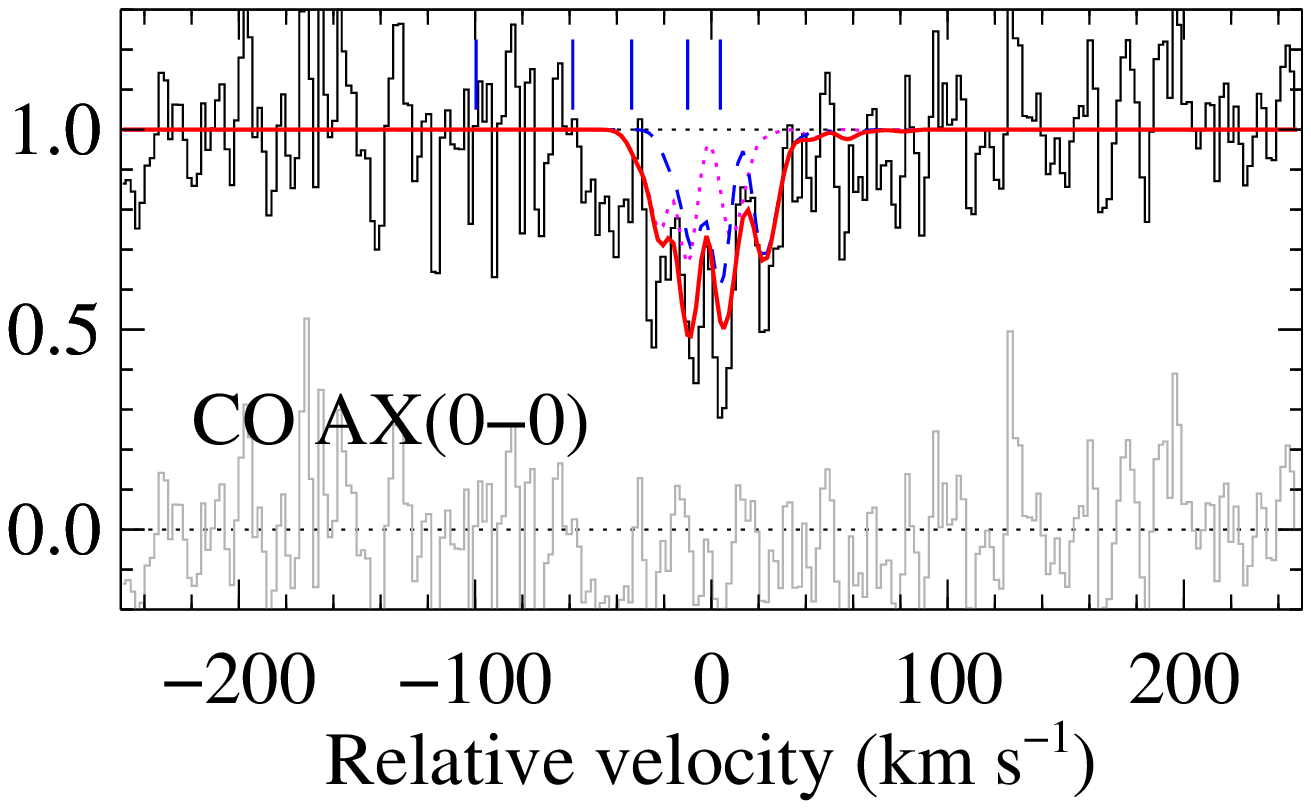}\\ 
\end{tabular}
\caption{\label{CO} Voigt profile fit to detected CO absorption lines. The origin of the velocity scale, 
is taken to be $z=1.6408$. Additional absorption in the CO A-X\,(1-0) panel at $v\sim150~\kms$ 
is due to the d-X\,(5-0) inter-band system. The contribution from individual components is shown at the 
position of the two reddest \SI\ components (short vertical marks) by dotted and dashed profiles.}
\end{figure}

Following \citet{Burgh07}, we use the excitation temperature as an external parameter. We fitted 
the CO profile using a IDL code based on MPFIT \citep{Markwardt09}, which performs 
$\chi^2$-minimisation by Levenberg-Marquardt technique.
Components associated to the reddest \SI\ components, at $z=1.64071$ and 1.64083, are clearly 
detected in the $J=0$ and $J=1$ rotational levels (see Fig.~\ref{CO}).
The S/N ratio of the data is not high enough so that an independent fit can be performed.
In addition, lines from different rotational levels are blended. We therefore fixed the redshifts 
and Doppler parameters to those obtained from \SI. 
We varied only the total CO column density for each of the two components, distributed 
among rotational levels up to $J=3$, using a single excitation temperature $T_{\rm ex}$. 

If we assume the excitation of CO is due to the Cosmic Microwave 
Background Radiation alone, i.e. $T_{\rm ex}=T_{\rm CMBR}(z=1.64)$, then the best fit model 
is achieved with $\chi^2_{\nu}=1.08$ for a total column density $\log N($CO$)=14.6$.
The model is superimposed on the observed spectrum in Fig.~\ref{CO}.

$T_{\rm ex}$ and $T_{\rm CMBR}$ are expected to be equal for low gas pressure \citep{Srianand08} 
and in the absence of UV pumping. Excitation by photon trapping becomes significant only when 
$N$(CO)$>10^{16}~\cmsq$ \citep{Burgh07}.
Unfortunately, unlike in the case of the $\zabs=2.42$ CO-bearing system 
towards \object{SDSS\,J143912$+$111740}, all observed neutral carbon lines are saturated 
and it is therefore impossible to estimate the pressure of the gas from the population of \CI\ 
fine-structure levels. 
We fitted different models along a grid of excitation temperatures and Doppler parameters. 
A good fit ($\chi^2 \la 1.2$) is achieved for a CO excitation temperature equals to or 
larger than the CMBR temperature and for $b>0.4$~\kms. We find $6<T_{\rm ex}<16$~K at the 
5\,$\sigma$ confidence level.

Assuming a single excitation temperature for all rotational levels could be 
considered a rough assumption. However, $T_{\rm 0J}$ is generally found constant for $J\le3$ 
in diffuse molecular gas in the local Universe \citep{Sheffer08}. Moreover, most of the optical 
depth of the CO profile is due to absorptions from $J=0$ and $J=1$. 
Actually, a two rotational levels model gives the same results.
We checked that the total CO column density only varies by less than 0.10~dex regardless 
of the $b$ and $T_{\rm ex}$ values as long as the condition $\chi^2<1.2$ is fulfilled.
The decomposition between the different components is quite uncertain however.

CO could be also present in the three bluest \SI\ components, 
although not detected directly at the 3\,$\sigma$ level. Because these 
components are far enough from the two detected CO components, they have 
very little influence on the excitation temperature. However, they could 
contribute to increase the overall CO column density 
(over the range $v=-150$ to $+$50~\kms) by 0.1~dex.

From the lower limit on $N(\CI)$ we derive $N($CO$)/N(\CI)\le 0.05$ which is typical
of what is measured in the diffuse molecular medium \citep{Federman80}.

\section{Dust and 2175~{\AA} UV bump}

The presence of dust grains influences the physical state of the gas through photo-electric 
heating, UV shielding, and formation of molecules on the surface of grains. 
It can be deduced from the depletion pattern of different elements 
(see Sect.~\ref{secmetals}) and/or from the reddening of the background QSO. 

\begin{figure}[!t]
\centering
\includegraphics[width=\hsize]{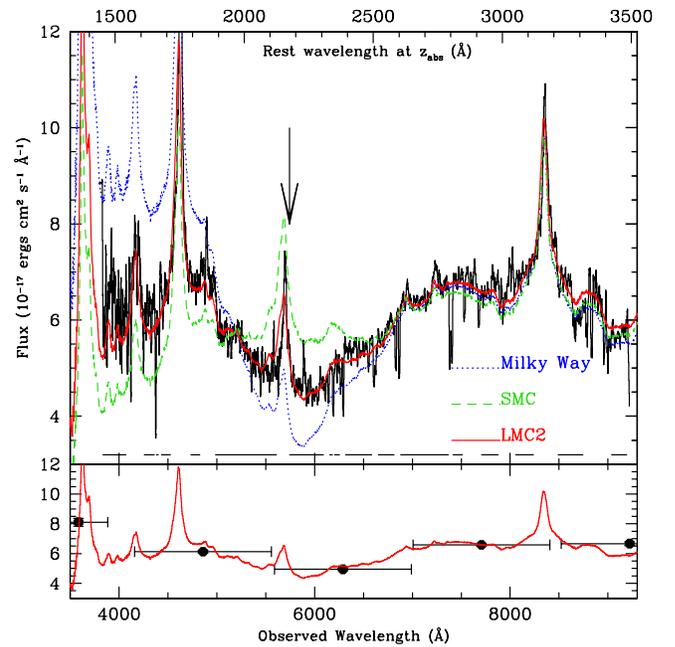}
\caption{SDSS spectrum of \thisqso\ fitted with the SDSS composite spectrum reddened 
by mean extinction curves from the Galaxy (dotted line), the LMC2 Supershell (solid line) and the SMC (dashed
line). The arrow indicates the position of the 2175~\angstrom\ feature at $\zabs=1.64$. 
The spectral regions used for $\chi^2$ minimisation are marked by solid lines
at the bottom of the top panel.
The observed spectrum is boxcar smoothed by 10 pixels for presentation purpose. The best fit and
the SDSS photometric points are shown in the bottom panel.
\label{bump}}
\end{figure}

The spectral energy distribution (SED) of \thisqsoshort\ is very red with u-K$\sim$4.5 mag. 
We note this value is only indicative as u (SDSS) and K (2MASS) magnitudes were measured with 
4 years interval and the exact u-K value could be different due to quasar optical variability.
The flux-calibrated SDSS spectrum of \thisqsoshort\ is visibly affected by 
reddening and also shows a clear curvature around 
2175~$\angstrom$ in the rest frame of the $z=1.64$ absorbing system (see Fig.~\ref{bump}). 
We investigate the possibility that the red colour of the quasar is due to the 
presence of dust in the CO-bearing system. 

We performed a $\chi^2$-minimisation between the data and a SDSS composite spectrum 
\citep{VandenBerk01} reddened by different types of extinction curves, namely that 
from the Small Magellanic Cloud (SMC), Large Magellanic Cloud Supershell (LMC2), 
Large Magellanic cloud (LMC) and the Galaxy (MW) \citep{Gordon03}. 
The strength of the 2175~\angstrom\ bump is very different from one extinction curve
to the other: while the UV bump is absent in the SMC extinction curve
it is strongest in the MW and LMC extinction curves. In addition, the SMC extinction curve
has a large UV extinction at $\lambda<2000~\angstrom$ compared to the LMC and MW extinction
curves. The average extinction curve of the LMC supershell (LMC2), which is part of the 30 Dor 
star forming region, has a bump strength and a UV extinction in-between the two extremes.

Results of our best fit models using different extinction curves are presented in 
Table~\ref{ext}. A detailed description of the 
procedure can be found in \citet{Srianand08bump}. 
The best reduced $\chi^2$ value is reached for the 
LMC2 Supershell dust extinction curve with E(B-V) = 0.27$\pm$0.02 for $R_{\rm V}$ = 2.7. 

\begin{table}
\caption{\label{ext} Results of fitting the SED of \thisqsoshort\ with different extinction curves}
\centering
\begin{tabular}{c c c c}
\hline
\hline
Extinction law &$R_V$ & E(B-V) & $\chi^2_{\nu}$\\
\hline
     MW    &3.1 &     $0.26\pm0.01$  &   3.0 \\
     LMC   &3.4 &     $0.24\pm0.02$  &   1.6 \\
     LMC2  &2.7 &     $0.27\pm0.02$  &   1.2 \\
     SMC   &2.7 &     $0.16\pm0.02$  &   2.1 \\
\hline
\end{tabular}\
\label{tabdust}
\end{table}

\begin{figure}[!t]
\centering
\includegraphics[bb=15 165 575 700, clip=, width=\hsize]{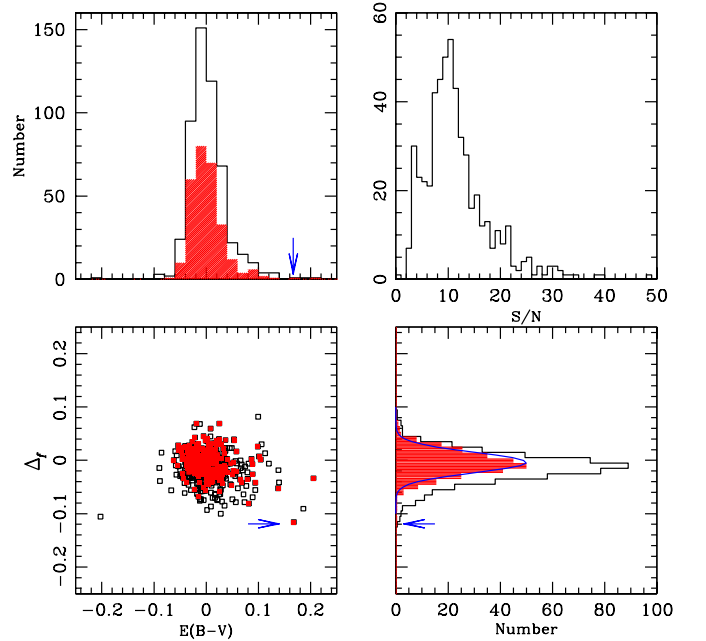}
\caption{Distributions of reddening (E(B-V)) and strength of the 2175~\angstrom\ bump 
($\Delta_f$) for the control sample of 529 quasars with $\zem\in[1.969,1.989]$. 
The arrows mark the position of \thisqso\ in the different panels.
Red filled squares and histograms indicate measurements in spectra with signal-to-noise ratios 
larger than 10. The top right panel shows the distribution of the signal-to-noise ratios.  
\label{control}}
\end{figure}

We performed a test on a sample of SDSS spectra to rule 
out the possibility that a peculiar intrinsic spectral shape is mistaken for a 
2175~$\angstrom$ bump \citep{Pitman00}. 
Our control sample consists of all 529 SDSS quasars with emission redshift 
within $\Delta z$=$\pm 0.01$ of that of \thisqsoshort\ ($\zem=1.979$). The distribution
of S/N ratios of the spectra\footnote{Median signal-to-noise ratio in the i-band (\texttt{SN\_I}) 
provided by SDSS together with the 1D spectra (spSpec*.fit)} is shown in the upper right panel of 
Fig.~\ref{control}. 
We fitted all quasar spectra using the SDSS composite spectrum reddened by a SMC extinction 
curve shifted to $\zabs=1.64$, {\sl which does not present any UV bump.}
Using $\chi^2$ minimisation we estimated, for each quasar, E(B-V) and the strength
of a possible 2175~\AA~feature. The latter is calculated as the integrated flux difference
between the fitted composite spectrum and the observed QSO spectrum in the region where the
2175~$\angstrom$ bump is redshifted (i.e. over the observed wavelength range, 
$5203 \le \lambda({\angstrom}) \le 6571$). We parametrised this quantity by 
$\Delta_f=\avg{F_{\rm QSO}/F_{\rm composite}-1}$. A plot of $\Delta_f$ versus E(B-V)
together with the distributions of the two parameters is presented in Fig.~\ref{control}. 
The positions of \thisqsoshort\ in these graphs are indicated by arrows.

As expected, QSOs with bad signal-to-noise ratio spectra (i.e. S/N $\le$ 10, open squares in the plots) 
are responsible for most of the scatter and a tail in the $\Delta_f$ distribution. 
Therefore, better confidence is achieved when selecting only high signal-to-noise spectra (i.e S/N $\ge$ 10).
Note \thisqsoshort\ has S/N~=~14.7.
From the scatter plot E(B-V) versus $\Delta_f$, it can be seen that the maximum deviation on 
both axis is seen for \thisqsoshort. The probability of getting E(B-V)~$>0.15$ is as low as 0.7\% when we consider 
only the high S/N ($> 10$) spectra. This corresponds to a 4.7~$\sigma$ significance if
we approximate the E(B-V) distribution for spectra of S/N~$>$~10 by a Gaussian function of mean zero 
and standard deviation 0.035.

There is only one quasar (\object{SDSS\,J131903$+$431034}) in the high S/N control sample 
with large E(B-V) in addition to \thisqsoshort. Interestingly, two strong \MgII\ systems are 
present in the spectrum of this QSO. Selecting quasar with red colours could therefore serve as a 
direct way to search for dusty intervening absorbers.
Concerning the 2175~$\angstrom$ absorption itself, only one quasar (the one presented here) has 
$\Delta_f<-0.10$. The corresponding probability is thus $0.4\%$. The mean $\Delta_f$ is zero with
a 1\,$\sigma$ dispersion of 0.035. So if we assume that the $\Delta f$ 
distribution for spectra with S/N~$>10$ is well modelled by a Gaussian distribution,  
the confidence level on the UV bump detection is 4.3\,$\sigma$.
The joint probability of measuring E(B-V)~$>$~0.15 and $\Delta_f$~$<-0.10$ by pure chance coincidence 
is $0.3$\%. 
In short, the exercise presented above rules out the possibility of the 2175~\angstrom\ structure being spurious 
at very high confidence level.

It is possible to estimate $N(\HI)$ from the reddening of the quasar, 
assuming the average LMC supershell relation between the column 
density of neutral gas and the extinction \citep{Gordon03}:
\begin{equation}
{N(\HI) \over {A_{\rm V}}}\sim (6.97\pm0.67)\times 10^{21}~\cmsq, 
\end{equation}
\noindent where $A_{\rm V}=R_{\rm V} \times E(B-V)$, and $R_{\rm V}$~=~2.76$\pm$0.09 is the average
value for the LMC2 supershell. 

We then get $N(\HI)\sim5\times10^{21}~\cmsq$. We can estimate a lower limit on $N$(\HI) using the
largest value of $A_{\rm v}/N(\HI)$~=~3.76$\times$10$^{-22}$~mag/(H cm$^{-2}$) measured in the 
LMC2 by \citet{Dobashi08}, $N(\HI) > 2\times10^{21}~\cmsq$. 
 We note that using E(B-V) values obtained with different extinction laws (Table~\ref{ext}) 
and the corresponding $A_{\rm V}/N(\HI)$ ratios provide very similar results. 
It would be very interesting to obtain a direct measurement of $N(\HI)$. 
However, the faintness of the quasar in the blue would make any determination 
with UVES very difficult.

\section{Discussion}

We have reported the detection of diffuse molecular gas at $\zabs=1.64$ 
towards \thisqsolong, following careful selection of the target among 
several thousand quasars from the SDSS. 

High spectral resolution observations with the Ultraviolet and Visual Echelle Spectrograph 
reveal the presence of CO molecules in the absorbing gas. This is only the second such detection 
in an intervening DLA along a quasar line of sight. From modelling the CO absorption with a multiple Voigt-profile 
pattern, we measure the CO column density to be $N$(CO)~$=4\times10^{14}\cmsq$ and an excitation 
temperature of the molecules, $6<T_{\rm ex}<16$~K at the 5\,$\sigma$ c.l., consistent with or higher 
than the expected temperature of the CMBR, $T_{\rm CMBR}=7.2~$K at $z=1.64$. 

The observed SED of the quasar is well reproduced if dust reddening happens 
in the CO absorber with an extinction law corresponding to that observed in the
LMC2 supershell including a 2175~\angstrom\ UV bump.
We derive significant reddening (E(B-V)=0.27) and a depletion pattern similar 
to what is seen in the diffuse molecular medium of the local interstellar medium. 

Statistical studies have shown that DLAs as an overall population are not 
producing significant reddening of the background quasar, with E(B-V) less than about 
0.04~mag \citep{Murphy04, Ellison05}. 
Moreover, \citet{Vladilo08} found a low mean extinction-to-gas ratio in DLAs
$A_{\rm V}/N$(\HI)$\sim$2-4$\times10^{-23}$~mag\,cm$^{2}$ in the SDSS-DR5. In addition, while 
strong \CaII\ absorbers seem to contain larger amounts of dust, they still produce only moderate 
reddening of the background quasar \citep[E(B-V)~$\sim0.1$;][]{Wild06}. Significant reddening, 
similar to that derived here, has only been found to date for a few individual absorption systems 
at low and intermediate redshift. \citet{Junkkarinen04} report E(B-V)~=~0.23 for the peculiar 
absorber at $\zabs=0.52$ towards AO\,0235$+$164 which also exhibits a 2175~{\AA} bump at the same 
redshift. E(B-V)~=~0.16 but no UV bump has been found toward SDSS\,J1323$-$0021 in a high-metallicity 
sub-DLA at $\zabs=0.72$ \citep{Khare04}. \citet{Ellison08b} detected DIBs in a \CaII\ absorber at 
$\zabs=0.16$ towards a reddened quasar with E(B-V)~=~0.23. More recently, \citet{Srianand08bump} 
measured E(B-V) about 0.3 and detected 2175~{\AA} dust features in two $z\sim1.3$ 21-cm absorbers.

It is interesting to note that the depletion pattern of the present system is similar to what has been observed in
the $\zabs = 2.418$ DLA towards SDSS\,J143912$+$111740 where CO is also detected 
\citep{Srianand08}. The present system has CO and \SI\ column 
densities six times larger than in the previous system when integrated column densities 
of \ZnII, \SiII\ and \FeII\ are only twice larger \citep{Noterdaeme08hd} .

The colours of \thisqsoshort\ make this object laying out of the 
quasar locus used by SDSS to select spectroscopic targets \citep{Richards02}.
More specifically, the object was flagged as 
\texttt{TARGET\_QSO\_REJECT} upon its colours and serendipitously assigned an excess 
spectroscopic fibre left over after the main samples of galaxies, LRGs, 
and quasars had been tiled \citep{Stoughton02}. We also note that this object is absent 
from the photometric catalogue of $\sim10^6$ quasars in SDSS-DR6 
\citep{Richards09}. Similarly, the dusty absorber at $z=1.3$ towards 
\object{SDSS\,J085244$+$343540} \citep{Srianand08bump}, is also absent from this catalogue and 
selected as SDSS spectroscopic target only because the background quasar is radio-loud.
This once again reiterates the fact that lines of sight towards colour-selected quasars 
are probably biased against the detection of the most relevant component of the interstellar 
medium, i.e., the diffuse molecular and translucent lines of sight.

It has been four decades since the first detections of the 2175~$\angstrom$ extinction 
feature in the spectra of hot stars \citep{Stecher65}. While it seems now clear that the 
presence of this bump is related to that of small dust grains, the composition of these latter 
remains debated. Theoretical and experimental studies have shown that carbon-rich organic
grains and polycyclic aromatic hydrocarbons (PAHs) are the most likely carriers 
of the 2175~\angstrom\ feature. Amorphous silicates, abundant in interplanetary dust particles 
could also contribute to the bump \citep{Bradley05}. 

The absence of UV bump in SMC lines-of-sight and in most star-burst galaxies lead \citet{Gordon97} 
to propose a trend in dust properties with star formation intensity.
Indeed, the UV bump carriers could be extremely sensitive to the local 
chemical enrichment and to the energy input in the gas \citep[e.g.][]{Whittet03, Gordon03, Noll07}. 
This is supported by theoretical work in the case of PAH clusters \citep{Rapacioli06}
and could explain why such feature is absent in the environment of most 
gamma-ray bursts \citep[see however][]{Eliasdottir09,Liang09} and Lyman break galaxies \citep{Vijh03}, 
and only present when the dust is shielded enough from the incident UV radiation.

An efficient shielding of the UV radiation field together with large metal abundances may 
thus be necessary for the presence of small carbon-rich grains \citep[see also][]{Sloan08}. 
These are the best conditions to efficiently form molecules 
\citep[e.g.][]{Ledoux03, Srianand05, Petitjean06, Noterdaeme08}, and in particular CO. 

The observations reported here demonstrate that carbonaceous molecules and grains already 
exist at $z=1.64$. This opens the exciting prospects of studying organic chemistry 
at high redshift.

\begin{acknowledgements}
We are grateful to the anonymous referee for insightful comments and suggestions.
PN acknowledges support from the french Ministry of Foreign and European Affairs. 
SL was supported by FONDECYT grant N$^{\rm o}1060823$.
We acknowledge the use of the Sloan Digital Sky Survey database (\url{http://www.sdss.org}). 

\end{acknowledgements}

\bibliographystyle{/scisoft/share/texmf/aa/aa-package/bibtex/aa}
\bibliography{../../mybib}

\begin{thebibliography}{54}
\expandafter\ifx\csname natexlab\endcsname\relax\def\natexlab#1{#1}\fi

\bibitem[{{Bradley} {et~al.}(2005){Bradley}, {Dai}, {Erni}, {Browning},
  {Graham}, {Weber}, {Smith}, {Hutcheon}, {Ishii}, {Bajt}, {Floss},
  {Stadermann}, \& {Sandford}}]{Bradley05}
{Bradley}, J., {Dai}, Z.~R., {Erni}, R., {et~al.} 2005, Science, 307, 244

\bibitem[{{Burgh} {et~al.}(2007){Burgh}, {France}, \& {McCandliss}}]{Burgh07}
{Burgh}, E.~B., {France}, K., \& {McCandliss}, S.~R. 2007, \apj, 658, 446

\bibitem[{{Dekker} {et~al.}(2000){Dekker}, {D'Odorico}, {Kaufer}, {Delabre}, \&
  {Kotzlowski}}]{Dekker00}
{Dekker}, H., {D'Odorico}, S., {Kaufer}, A., {Delabre}, B., \& {Kotzlowski}, H.
  2000, in Proc. SPIE Vol. 4008, p. 534-545, Optical and IR Telescope
  Instrumentation and Detectors, Masanori Iye; Alan F. Moorwood; Eds., 534--545

\bibitem[{{Dobashi} {et~al.}(2008){Dobashi}, {Bernard}, {Hughes}, {Paradis},
  {Reach}, \& {Kawamura}}]{Dobashi08}
{Dobashi}, K., {Bernard}, J.-P., {Hughes}, A., {et~al.} 2008, \aap, 484, 205

\bibitem[{{Eidelsberg} \& {Rostas}(2003)}]{Eidelsberg03}
{Eidelsberg}, M. \& {Rostas}, F. 2003, \apjs, 145, 89

\bibitem[{{El{\'{\i}}asd{\'o}ttir} {et~al.}(2009){El{\'{\i}}asd{\'o}ttir},
  {Fynbo}, {Hjorth}, {Ledoux}, {Watson}, {Andersen}, {Malesani}, {Vreeswijk},
  {Prochaska}, {Sollerman}, \& {Jaunsen}}]{Eliasdottir09}
{El{\'{\i}}asd{\'o}ttir}, {\'A}., {Fynbo}, J.~P.~U., {Hjorth}, J., {et~al.}
  2009, \apj, 697, 1725

\bibitem[{{Ellison} {et~al.}(2005){Ellison}, {Hall}, \& {Lira}}]{Ellison05}
{Ellison}, S.~L., {Hall}, P.~B., \& {Lira}, P. 2005, \aj, 130, 1345

\bibitem[{{Ellison} {et~al.}(2008){Ellison}, {York}, {Murphy}, {Zych}, {Smith},
  \& {Sarre}}]{Ellison08b}
{Ellison}, S.~L., {York}, B.~A., {Murphy}, M.~T., {et~al.} 2008, \mnras, 383,
  L30

\bibitem[{{Federman} {et~al.}(1980){Federman}, {Glassgold}, {Jenkins}, \&
  {Shaya}}]{Federman80}
{Federman}, S.~R., {Glassgold}, A.~E., {Jenkins}, E.~B., \& {Shaya}, E.~J.
  1980, \apj, 242, 545

\bibitem[{{Fontana} \& {Ballester}(1995)}]{Fontana95}
{Fontana}, A. \& {Ballester}, P. 1995, The Messenger, 80, 37

\bibitem[{{Gordon} {et~al.}(1997){Gordon}, {Calzetti}, \& {Witt}}]{Gordon97}
{Gordon}, K.~D., {Calzetti}, D., \& {Witt}, A.~N. 1997, \apj, 487, 625

\bibitem[{{Gordon} {et~al.}(2003){Gordon}, {Clayton}, {Misselt}, {Landolt}, \&
  {Wolff}}]{Gordon03}
{Gordon}, K.~D., {Clayton}, G.~C., {Misselt}, K.~A., {Landolt}, A.~U., \&
  {Wolff}, M.~J. 2003, \apj, 594, 279

\bibitem[{{Grevesse} {et~al.}(2007){Grevesse}, {Asplund}, \&
  {Sauval}}]{Grevesse07}
{Grevesse}, N., {Asplund}, M., \& {Sauval}, A.~J. 2007, Space Science Reviews,
  130, 105

\bibitem[{{Junkkarinen} {et~al.}(2004){Junkkarinen}, {Cohen}, {Beaver},
  {Burbidge}, {Lyons}, \& {Madejski}}]{Junkkarinen04}
{Junkkarinen}, V.~T., {Cohen}, R.~D., {Beaver}, E.~A., {et~al.} 2004, \apj,
  614, 658

\bibitem[{{Khare} {et~al.}(2004){Khare}, {Kulkarni}, {Lauroesch}, {York},
  {Crotts}, \& {Nakamura}}]{Khare04}
{Khare}, P., {Kulkarni}, V.~P., {Lauroesch}, J.~T., {et~al.} 2004, \apj, 616,
  86

\bibitem[{{Ledoux} {et~al.}(2003){Ledoux}, {Petitjean}, \&
  {Srianand}}]{Ledoux03}
{Ledoux}, C., {Petitjean}, P., \& {Srianand}, R. 2003, \mnras, 346, 209

\bibitem[{{Ledoux} {et~al.}(2002){Ledoux}, {Srianand}, \&
  {Petitjean}}]{Ledoux02}
{Ledoux}, C., {Srianand}, R., \& {Petitjean}, P. 2002, \aap, 392, 781

\bibitem[{{Liang} \& {Li}(2009)}]{Liang09}
{Liang}, S.~L. \& {Li}, A. 2009, \apjl, 690, L56

\bibitem[{{Markwardt}(2009)}]{Markwardt09}
{Markwardt}, C.~B. 2009, ArXiv e-prints 0902.2850

\bibitem[{{Morton} \& {Noreau}(1994)}]{Morton94}
{Morton}, D.~C. \& {Noreau}, L. 1994, \apjs, 95, 301

\bibitem[{{Motta} {et~al.}(2002){Motta}, {Mediavilla}, {Mu{\~n}oz}, {Falco},
  {Kochanek}, {Arribas}, {Garc{\'{\i}}a-Lorenzo}, {Oscoz}, \&
  {Serra-Ricart}}]{Motta02}
{Motta}, V., {Mediavilla}, E., {Mu{\~n}oz}, J.~A., {et~al.} 2002, \apj, 574,
  719

\bibitem[{{Murphy} \& {Liske}(2004)}]{Murphy04}
{Murphy}, M.~T. \& {Liske}, J. 2004, \mnras, 354, L31

\bibitem[{{Noll} {et~al.}(2007){Noll}, {Pierini}, {Pannella}, \&
  {Savaglio}}]{Noll07}
{Noll}, S., {Pierini}, D., {Pannella}, M., \& {Savaglio}, S. 2007, \aap, 472,
  455

\bibitem[{{Noterdaeme} {et~al.}(2008{\natexlab{a}}){Noterdaeme}, {Ledoux},
  {Petitjean}, \& {Srianand}}]{Noterdaeme08}
{Noterdaeme}, P., {Ledoux}, C., {Petitjean}, P., \& {Srianand}, R.
  2008{\natexlab{a}}, \aap, 481, 327

\bibitem[{{Noterdaeme} {et~al.}(2008{\natexlab{b}}){Noterdaeme}, {Petitjean},
  {Ledoux}, {Srianand}, \& {Ivanchik}}]{Noterdaeme08hd}
{Noterdaeme}, P., {Petitjean}, P., {Ledoux}, C., {Srianand}, R., \& {Ivanchik},
  A. 2008{\natexlab{b}}, \aap, 491, 397

\bibitem[{{Petitjean} {et~al.}(1992){Petitjean}, {Bergeron}, \&
  {Puget}}]{Petitjean92}
{Petitjean}, P., {Bergeron}, J., \& {Puget}, J.~L. 1992, \aap, 265, 375

\bibitem[{{Petitjean} {et~al.}(2006){Petitjean}, {Ledoux}, {Noterdaeme}, \&
  {Srianand}}]{Petitjean06}
{Petitjean}, P., {Ledoux}, C., {Noterdaeme}, P., \& {Srianand}, R. 2006, \aap,
  456, L9

\bibitem[{{Petitjean} {et~al.}(2000){Petitjean}, {Srianand}, \&
  {Ledoux}}]{Petitjean00}
{Petitjean}, P., {Srianand}, R., \& {Ledoux}, C. 2000, \aap, 364, L26

\bibitem[{{Pettini} {et~al.}(1997){Pettini}, {Smith}, {King}, \&
  {Hunstead}}]{Pettini97}
{Pettini}, M., {Smith}, L.~J., {King}, D.~L., \& {Hunstead}, R.~W. 1997, \apj,
  486, 665

\bibitem[{{Pitman} {et~al.}(2000){Pitman}, {Clayton}, \& {Gordon}}]{Pitman00}
{Pitman}, K.~M., {Clayton}, G.~C., \& {Gordon}, K.~D. 2000, \pasp, 112, 537

\bibitem[{{Prochaska} {et~al.}(2009){Prochaska}, {Sheffer}, {Perley}, {Bloom},
  {Lopez}, {Dessauges-Zavadsky}, {Chen}, {Filippenko}, {Ganeshalingam}, {Li},
  {Miller}, \& {Starr}}]{Prochaska09}
{Prochaska}, J.~X., {Sheffer}, Y., {Perley}, D.~A., {et~al.} 2009, \apjl, 691,
  L27

\bibitem[{{Prochaska} \& {Wolfe}(2002)}]{Prochaska02}
{Prochaska}, J.~X. \& {Wolfe}, A.~M. 2002, \apj, 566, 68

\bibitem[{{Quast} {et~al.}(2008){Quast}, {Reimers}, \& {Baade}}]{Quast08}
{Quast}, R., {Reimers}, D., \& {Baade}, R. 2008, \aap, 477, 443

\bibitem[{{Rapacioli} {et~al.}(2006){Rapacioli}, {Calvo}, {Joblin}, {Parneix},
  {Toublanc}, \& {Spiegelman}}]{Rapacioli06}
{Rapacioli}, M., {Calvo}, F., {Joblin}, C., {et~al.} 2006, \aap, 460, 519

\bibitem[{{Richards} {et~al.}(2002){Richards}, {Fan}, {Newberg}, {Strauss},
  {Vanden Berk}, {Schneider}, {Yanny}, {Boucher}, {Burles}, {Frieman}, {Gunn},
  {Hall}, {Ivezi{\'c}}, {Kent}, {Loveday}, {Lupton}, {Rockosi}, {Schlegel},
  {Stoughton}, {SubbaRao}, \& {York}}]{Richards02}
{Richards}, G.~T., {Fan}, X., {Newberg}, H.~J., {et~al.} 2002, \aj, 123, 2945

\bibitem[{{Richards} {et~al.}(2009){Richards}, {Myers}, {Gray}, {Riegel},
  {Nichol}, {Brunner}, {Szalay}, {Schneider}, \& {Anderson}}]{Richards09}
{Richards}, G.~T., {Myers}, A.~D., {Gray}, A.~G., {et~al.} 2009, \apjs, 180, 67

\bibitem[{{Sheffer} {et~al.}(2008){Sheffer}, {Rogers}, {Federman}, {Abel},
  {Gredel}, {Lambert}, \& {Shaw}}]{Sheffer08}
{Sheffer}, Y., {Rogers}, M., {Federman}, S.~R., {et~al.} 2008, \apj, 687, 1075

\bibitem[{{Sloan} {et~al.}(2008){Sloan}, {Kraemer}, {Wood}, {Zijlstra},
  {Bernard-Salas}, {Devost}, \& {Houck}}]{Sloan08}
{Sloan}, G.~C., {Kraemer}, K.~E., {Wood}, P.~R., {et~al.} 2008, \apj, 686, 1056

\bibitem[{{Srianand} {et~al.}(2008{\natexlab{a}}){Srianand}, {Gupta},
  {Petitjean}, {Noterdaeme}, \& {Saikia}}]{Srianand08bump}
{Srianand}, R., {Gupta}, N., {Petitjean}, P., {Noterdaeme}, P., \& {Saikia},
  D.~J. 2008{\natexlab{a}}, \mnras, 391, L69

\bibitem[{{Srianand} {et~al.}(2008{\natexlab{b}}){Srianand}, {Noterdaeme},
  {Ledoux}, \& {Petitjean}}]{Srianand08}
{Srianand}, R., {Noterdaeme}, P., {Ledoux}, C., \& {Petitjean}, P.
  2008{\natexlab{b}}, \aap, 482, L39

\bibitem[{{Srianand} {et~al.}(2005){Srianand}, {Petitjean}, {Ledoux},
  {Ferland}, \& {Shaw}}]{Srianand05}
{Srianand}, R., {Petitjean}, P., {Ledoux}, C., {Ferland}, G., \& {Shaw}, G.
  2005, \mnras, 362, 549

\bibitem[{{Stecher}(1965)}]{Stecher65}
{Stecher}, T.~P. 1965, \apj, 142, 1683

\bibitem[{{Stoughton} {et~al.}(2002){Stoughton}, {Lupton}, {Bernardi},
  {Blanton}, {Burles}, {Castander}, {Connolly}, {Eisenstein}, {Frieman},
  {Hennessy}, {Hindsley}, {Ivezi{\'c}}, {Kent}, {Kunszt}, {Lee}, {Meiksin},
  {Munn}, {Newberg}, {Nichol}, {Nicinski}, {Pier}, {Richards}, {Richmond},
  {Schlegel}, {Smith}, {Strauss}, {SubbaRao}, {Szalay}, {Thakar}, {Tucker},
  {Vanden Berk}, {Yanny}, {Adelman}, {Anderson}, {Anderson}, {Annis},
  {Bahcall}, {Bakken}, {Bartelmann}, {Bastian}, {Bauer}, {Berman},
  {B{\"o}hringer}, {Boroski}, {Bracker}, {Briegel}, {Briggs}, {Brinkmann},
  {Brunner}, {Carey}, {Carr}, {Chen}, {Christian}, {Colestock}, {Crocker},
  {Csabai}, {Czarapata}, {Dalcanton}, {Davidsen}, {Davis}, {Dehnen},
  {Dodelson}, {Doi}, {Dombeck}, {Donahue}, {Ellman}, {Elms}, {Evans}, {Eyer},
  {Fan}, {Federwitz}, {Friedman}, {Fukugita}, {Gal}, {Gillespie}, {Glazebrook},
  {Gray}, {Grebel}, {Greenawalt}, {Greene}, {Gunn}, {de Haas}, {Haiman},
  {Haldeman}, {Hall}, {Hamabe}, {Hansen}, {Harris}, {Harris}, {Harvanek},
  {Hawley}, {Hayes}, {Heckman}, {Helmi}, {Henden}, {Hogan}, {Hogg}, {Holmgren},
  {Holtzman}, {Huang}, {Hull}, {Ichikawa}, {Ichikawa}, {Johnston}, {Kauffmann},
  {Kim}, {Kimball}, {Kinney}, {Klaene}, {Kleinman}, {Klypin}, {Knapp},
  {Korienek}, {Krolik}, {Kron}, {Krzesi{\'n}ski}, {Lamb}, {Leger},
  {Limmongkol}, {Lindenmeyer}, {Long}, {Loomis}, {Loveday}, {MacKinnon},
  {Mannery}, {Mantsch}, {Margon}, {McGehee}, {McKay}, {McLean}, {Menou},
  {Merelli}, {Mo}, {Monet}, {Nakamura}, {Narayanan}, {Nash}, {Neilsen},
  {Newman}, {Nitta}, {Odenkirchen}, {Okada}, {Okamura}, {Ostriker}, {Owen},
  {Pauls}, {Peoples}, {Peterson}, {Petravick}, {Pope}, {Pordes}, {Postman},
  {Prosapio}, {Quinn}, {Rechenmacher}, {Rivetta}, {Rix}, {Rockosi}, {Rosner},
  {Ruthmansdorfer}, {Sandford}, {Schneider}, {Scranton}, {Sekiguchi}, {Sergey},
  {Sheth}, {Shimasaku}, {Smee}, {Snedden}, {Stebbins}, {Stubbs}, {Szapudi},
  {Szkody}, {Szokoly}, {Tabachnik}, {Tsvetanov}, {Uomoto}, {Vogeley}, {Voges},
  {Waddell}, {Walterbos}, {Wang}, {Watanabe}, {Weinberg}, {White}, {White},
  {Wilhite}, {Wolfe}, {Yasuda}, {York}, {Zehavi}, \& {Zheng}}]{Stoughton02}
{Stoughton}, C., {Lupton}, R.~H., {Bernardi}, M., {et~al.} 2002, \aj, 123, 485

\bibitem[{{Tumlinson} {et~al.}(2007){Tumlinson}, {Prochaska}, {Chen},
  {Dessauges-Zavadsky}, \& {Bloom}}]{Tumlinson07}
{Tumlinson}, J., {Prochaska}, J.~X., {Chen}, H.-W., {Dessauges-Zavadsky}, M.,
  \& {Bloom}, J.~S. 2007, \apj, 668, 667

\bibitem[{{Vanden Berk} {et~al.}(2001){Vanden Berk}, {Richards}, {Bauer},
  {Strauss}, {Schneider}, {Heckman}, {York}, {Hall}, {Fan}, {Knapp},
  {Anderson}, {Annis}, {Bahcall}, {Bernardi}, {Briggs}, {Brinkmann}, {Brunner},
  {Burles}, {Carey}, {Castander}, {Connolly}, {Crocker}, {Csabai}, {Doi},
  {Finkbeiner}, {Friedman}, {Frieman}, {Fukugita}, {Gunn}, {Hennessy},
  {Ivezi{\'c}}, {Kent}, {Kunszt}, {Lamb}, {Leger}, {Long}, {Loveday}, {Lupton},
  {Meiksin}, {Merelli}, {Munn}, {Newberg}, {Newcomb}, {Nichol}, {Owen}, {Pier},
  {Pope}, {Rockosi}, {Schlegel}, {Siegmund}, {Smee}, {Snir}, {Stoughton},
  {Stubbs}, {SubbaRao}, {Szalay}, {Szokoly}, {Tremonti}, {Uomoto}, {Waddell},
  {Yanny}, \& {Zheng}}]{VandenBerk01}
{Vanden Berk}, D.~E., {Richards}, G.~T., {Bauer}, A., {et~al.} 2001, \aj, 122,
  549

\bibitem[{{Vijh} {et~al.}(2003){Vijh}, {Witt}, \& {Gordon}}]{Vijh03}
{Vijh}, U.~P., {Witt}, A.~N., \& {Gordon}, K.~D. 2003, \apj, 587, 533

\bibitem[{{Vladilo} {et~al.}(2008){Vladilo}, {Prochaska}, \&
  {Wolfe}}]{Vladilo08}
{Vladilo}, G., {Prochaska}, J.~X., \& {Wolfe}, A.~M. 2008, \aap, 478, 701

\bibitem[{{Vreeswijk} {et~al.}(2007){Vreeswijk}, {Ledoux}, {Smette}, {Ellison},
  {Jaunsen}, {Andersen}, {Fruchter}, {Fynbo}, {Hjorth}, {Kaufer}, {M{\o}ller},
  {Petitjean}, {Savaglio}, \& {Wijers}}]{Vreeswijk07}
{Vreeswijk}, P.~M., {Ledoux}, C., {Smette}, A., {et~al.} 2007, \aap, 468, 83

\bibitem[{{Welty} {et~al.}(1999){Welty}, {Frisch}, {Sonneborn}, \&
  {York}}]{Welty99}
{Welty}, D.~E., {Frisch}, P.~C., {Sonneborn}, G., \& {York}, D.~G. 1999, \apj,
  512, 636

\bibitem[{{Whittet}(2003)}]{Whittet03}
{Whittet}, D.~C.~B., ed. 2003, {Dust in the galactic environment}

\bibitem[{{Wild} {et~al.}(2006){Wild}, {Hewett}, \& {Pettini}}]{Wild06}
{Wild}, V., {Hewett}, P.~C., \& {Pettini}, M. 2006, \mnras, 367, 211

\bibitem[{{Wolfe} {et~al.}(2005){Wolfe}, {Gawiser}, \& {Prochaska}}]{Wolfe05}
{Wolfe}, A.~M., {Gawiser}, E., \& {Prochaska}, J.~X. 2005, \araa, 43, 861

\bibitem[{{Wucknitz} {et~al.}(2003){Wucknitz}, {Wisotzki}, {Lopez}, \&
  {Gregg}}]{Wucknitz03}
{Wucknitz}, O., {Wisotzki}, L., {Lopez}, S., \& {Gregg}, M.~D. 2003, \aap, 405,
  445

\bibitem[{{Zwaan} \& {Prochaska}(2006)}]{Zwaan06}
{Zwaan}, M.~A. \& {Prochaska}, J.~X. 2006, \apj, 643, 675

\end{thebibliography}
\end{document}